\documentclass[11pt]{article}
\usepackage{jheppub}
\usepackage{amsmath,amssymb,amsfonts,graphicx,slashed,amsthm,mathtools,upgreek, enumerate, tensor, subfig}
\usepackage[dvipsnames]{xcolor}
\usepackage{arydshln}
\usepackage{slashed}
\usepackage{listings}
\usepackage{comment}
\usepackage{hyperref}
\usepackage[utf8]{inputenc}
\usepackage[titletoc]{appendix}

\makeatletter
\def\@fpheader{\relax}
\makeatother

\DeclareMathOperator{\Tr}{Tr}

\newcommand{\avg}[1]{\langle #1 \rangle}
\newcommand\be{\begin{equation}}
\newcommand\ee{\end{equation}}
\newcommand\bea{\begin{eqnarray}}
\newcommand\eea{\end{eqnarray}}

\newcommand\eref[1]{(\ref{#1})}
\newcommand\bc{\begin{center}}
\newcommand\ec{\end{center}}

\numberwithin{equation}{section} 
\allowdisplaybreaks

\title{Disentangling a Deep Learned Volume Formula}

\author[a]{Jessica Craven}
\author[a]{\!, Vishnu Jejjala}
\author[b]{\!, Arjun Kar}

\affiliation[\,a]{Mandelstam Institute for Theoretical Physics, School of Physics, NITheP, and CoE-MaSS,\\
University of the Witwatersrand, Johannesburg, WITS 2050, South Africa}
\affiliation[\,b]{Department of Physics and Astronomy, University of British Columbia,\\
6224 Agricultural Road, Vancouver, BC V6T 1Z1, Canada}

\emailAdd{jessica.craven1@students.wits.ac.za}
\emailAdd{vishnu@neo.phys.wits.ac.za}
\emailAdd{arjunkar@phas.ubc.ca}

\abstract{
We present a simple phenomenological formula which approximates the hyperbolic volume of a knot using only a single evaluation of its Jones polynomial at a root of unity.
The average error is just $2.86$\% on the first $1.7$ million knots, which represents a large improvement over previous formulas of this kind.
To find the approximation formula, we use layer-wise relevance propagation to reverse engineer a black box neural network which achieves a similar average error for the same approximation task when trained on $10$\% of the total dataset.
The particular roots of unity which appear in our analysis cannot be written as $e^{2\pi i / (k+2)}$ with integer $k$; therefore, the relevant Jones polynomial evaluations are not given by unknot-normalized expectation values of Wilson loop operators in conventional $SU(2)$ Chern--Simons theory with level $k$.
Instead, they correspond to an analytic continuation of such expectation values to fractional level.
We briefly review the continuation procedure and comment on the presence of certain Lefschetz thimbles, to which our approximation formula is sensitive, in the analytically continued Chern--Simons integration cycle.
}

\begin{document}

\maketitle
\parskip=10pt

\section{Introduction}
The intersection of mathematical phenomenology and physical mathematics has historically led to profound insights into mathematical structures and the natural world.
A modern example is the development of mirror symmetry, first as an observation about the non-uniqueness of Calabi--Yau compactifications of superstring theory~\cite{Dixon:1987bg,Lerche:1989uy}, then as a statement about mirror pairs of Calabi--Yau manifolds with Hodge numbers interchanged~\cite{Candelas:1989hd} and as a relationship between Gepner models~\cite{Greene:1990ud}.
These dualities among physical theories and the success of mirror symmetry as a tool to solve problems in enumerative geometry~\cite{Candelas:1990rm} supplied an invitation and a motivation for later rigorous mathematical theorems~\cite{giv,Lian:1999rn}.

Knot theory presents another setting in which physics (quantum field theory) provides a novel window into mathematics.
A knot is an embedding of $S^1 \subset S^3$.
Because the same knot can be drawn in multiple ways, topological invariants provide labels for identifying a given knot $K$.
Two of these topological invariants are the Jones polynomial, $J_2(K;q)$~\cite{Jones1987}, and, for hyperbolic knots, the volume of the knot complement, $\text{Vol}(S^3 \setminus K)$~\cite{thu}.
The Jones polynomial, as we review below, has a physical realization in terms of Wilson loop observables in Chern--Simons gauge theory~\cite{Witten:1988hf}, where it can be generalized to the colored Jones polynomial $J_R(K;q)$, with $R$ a representation of the gauge group.
These two topological invariants are related through the volume conjecture~\cite{Kashaev1997,Murakami2001,Gukov:2003na}:
\be
\lim_{n\to\infty} \frac{2\pi \log|J_{n}(K;\omega_n)|}{n} = \text{Vol}(S^3 \setminus K) ~, \label{eq:vc}
\ee
where $K$ is a hyperbolic knot, the color $n$ denotes the $n$-dimensional irreducible representation of the gauge group $SU(2)$ in which the trace of the Wilson loop is taken, and $\omega_n = e^{2\pi i /n}$.
The ordinary Jones polynomial corresponds to taking the fundamental representation $n=2$.

While the large $n$ limit of the colored Jones polynomial takes center stage in the volume conjecture, it turns out that the ordinary Jones polynomial also conveys some information about the volume.
Dunfield initially observed the trend that, for alternating knots up to $13$ crossings, a simple linear function of $\log |J_2(K;-1)|$ was approximately proportional to $\text{Vol}(S^3 \setminus K)$~\cite{Dunfield2000}.
As noted already in that work, this relationship is mysterious because the coefficients of the linear function are not what one would expect by simply writing down the expression on the left hand side of~\eqref{eq:vc} with $n=2$.
So, the linear relationship cannot be explained by suggesting that the volume conjecture converges quickly.
Indeed, it is known that the left hand side of~\eqref{eq:vc} is not even monotonic in $n$ for certain knots, and~\cite{Garoufalidis_2005} conjectured that convergence is only eventually monotonic.

Subsequently, a neural network predicted the hyperbolic volume from the Jones polynomial with $97.55\pm 0.10$\% accuracy for all hyperbolic knots up to $15$ crossings using only $10$\% of the dataset of $313,209$ knots for training~\cite{Jejjala:2019kio}.
The input to the network was a vector consisting of the maximum and minimum degrees of the Jones polynomial along with its integer coefficients.
The network's accuracy is essentially optimal for the following reason:
knots with different hyperbolic volumes can have the same Jones polynomial, and when this happens, the volumes typically differ by about $3$\%.
This represented a large improvement over the approximation formula using $J_2(K;-1)$, but introduced many more free parameters (the weights and biases of the neural network) and completely obscured the true functional dependence on $J_2(K;q)$.

This work in experimental mathematics established that there is more information about the volume contained in the full Jones polynomial than in the single evaluation $J_2(K;-1)$.\footnote{We refer loosely to the information contained in certain invariants, as all of our results are numerical and therefore ``probably approximately correct''~\cite{Valiant1984}.}
Indeed, training on only $319$ knots ($0.1$\% of the dataset) is sufficient to extract this additional information.
Moreover, the predictions apply to all hyperbolic knots, not just the alternating ones.
The neural network, however, is a black box that learns semantics without knowing any syntax.
While machine learning successfully identifies relationships between topological invariants of hyperbolic knots, we do not have an analytic understanding of why this occurs or how the machine learns the relationship in the first place.
The aim of this paper is to determine which aspects of the input are most important for the neural network's considerations and to use this to deduce a simpler functional relationship between the Jones polynomial and the hyperbolic volume that depends on only an $O(1)$ number of free parameters.
In other words, we seek an approximation formula which similarly outperforms the formula based on $J_2(K;-1)$, but which does not rely on the complicated structure of a neural network.

The main result obtained in this work is an approximation formula for the hyperbolic volume of a knot in terms of a single evaluation of its Jones polynomial.
\begin{equation}
    V_{3/4}(K) = 6.20 \log{(|J_2(K;e^{3\pi i/4})| + 6.77)} - 0.94 ~.
\label{eq:result}
\end{equation}
This formula is numerical in nature, and we have no sharp error bounds, but it achieves an accuracy of more than $97$\% on the first $1.7$ million hyperbolic knots.
The phase $e^{3\pi i/ 4}$ can be adjusted to some degree to obtain similar formulas which perform almost as well; we explore these alternatives.
We obtain~\eqref{eq:result} and its analogues by reverse engineering the behavior of the aforementioned neural network, a task which is generally considered difficult in the machine learning community.
In our case, it is possible due to the simplicity of the underlying neural network architecture and the power of layer-wise relevance propagation, a network analysis technique which we review.

Since the phase $e^{3\pi i / 4}$ cannot be written as $e^{2\pi i / (k+2)}$ for integer $k$,~\eqref{eq:result} suggests that we should consider analytically continued Chern--Simons theory, which was previously explored as a possible route to understand the volume conjecture~\cite{Witten:2010cx}.
Specifically, the evaluations of the Jones polynomial that are relevant for us correspond to fractional Chern--Simons levels $k=\frac23$ and $k=\frac12$, which must be understood in the context of the analytically continued theory due to the integer level restriction normally enforced by gauge invariance.
We provide a review of the main points of the analytically continued theory, as our results have some speculative bearing on which Lefschetz thimbles should contribute to the path integral at various values of $n$ and $k$.
Our interpretations pass several sanity checks when compared with the main lessons of~\cite{Witten:2010cx}, and we use our results to formulate an alternative version of the volume conjecture.

\begin{figure}[t]
    \centering
    \includegraphics[scale=.8]{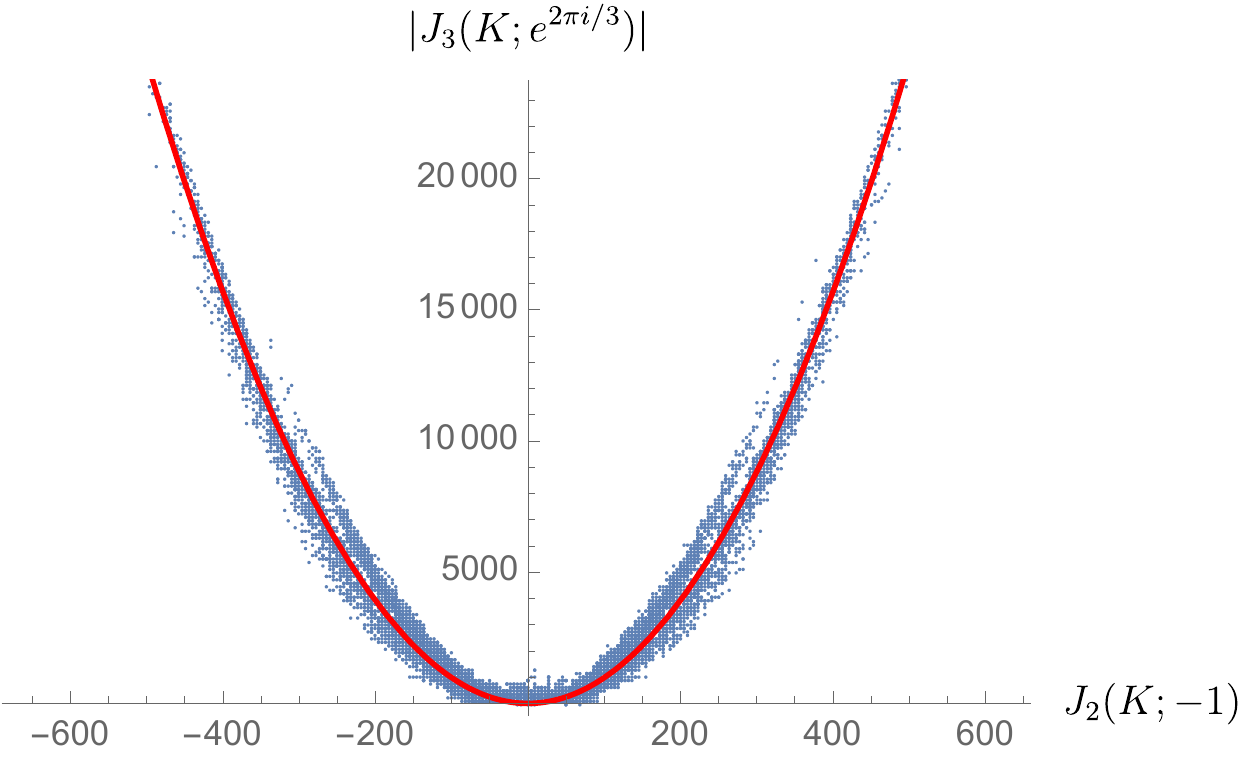}
    \caption{The $n=2$ and $n=3$ colored Jones polynomials evaluated at roots of unity show a quadratic relation.}
    \label{fig:j2j3}
\end{figure}

Though we do not have a complete explanation for why~\eqref{eq:result} predicts the volume so well, its implications are intriguing.
It points to the existence of a sort of quantum/semiclassical duality between $SU(2)$ and $SL(2,\mathbb{C})$ Chern--Simons theory, since some simple numerical coefficients are enough to transform a strong coupling object (the Jones polynomial at small $k$) into a weak coupling one (the hyperbolic volume).
This is reminiscent of the shift $k \to k+2$ induced by the one-loop correction in the $SU(2)$ Chern--Simons path integral~\cite{Witten:1988hf}, which transforms a semiclassical approximation into a more quantum result via a single $O(1)$ parameter.
We can explore this phenomenon in a preliminary way by checking whether $J_2$ contains any information about $J_n$ for $n>2$ and some $O(1)$ phases like the ones appearing in the volume conjecture.
Computing $11,921$ colored Jones polynomials in the adjoint representation of $SU(2)$, we notice from Figure~\ref{fig:j2j3} that
\be
|J_3(K;\omega_3)| \sim J_2(K;\omega_2)^2 ~.
\ee
Comparing to the volume conjecture~\eref{eq:vc}, this suggests that the ordinary Jones polynomial has some knowledge of the behavior of higher colors and in particular the $n\to \infty$ limit.\footnote{
This analysis was performed in collaboration with Onkar Parrikar.}

The organization of this paper is as follows.
In Section~\ref{sec:knots}, we review relevant aspects of Chern--Simons theory, including its relationship to knot invariants and its analytic continuation.
In Section~\ref{sec:ml}, we describe the machine learning methods we employ, particularly neural networks and layer-wise relevance propagation.
In Section~\ref{sec:deep}, we analyze the structure of the neural network and, using layer-wise relevance propagation,  deduce a relationship between evaluations of the Jones polynomial of a hyperbolic knot and the volume of its knot complement.
We also comment on the implications of our numerics for the presence of certain integration cycles in the analytically continued Chern--Simons path integral as a function of coupling.
In Section~\ref{sec:disc}, we discuss the results we obtain and propose some future directions.
We also provide several appendices which outline details about the implementation of our machine learning algorithms (Appendix~\ref{sec:nnsetup}), results relating the scaling of the Jones polynomial coefficients with the hyperbolic volume (Appendix~\ref{sec:khovanov}), details concerning the various normalizations of the Jones polynomial in the mathematics and physics literature (Appendix~\ref{sec:norm}), a data analysis of knot invariants using t-distributed stochastic neighbor embedding (Appendix~\ref{sec:tsne}), and finally, an overview of related experiments (Appendix~\ref{sec:more_exp}).

\section{Chern--Simons theory}\label{sec:knots}

We review several aspects of Chern--Simons theory, its relation to the colored Jones polynomials and hyperbolic volumes of knots, and its analytic continuation away from integer level.

\subsection{Knot invariants}

Chern--Simons theory is a three-dimensional topological quantum field theory which provides a unifying language for the knot invariants with which we will be concerned~\cite{Witten:1988hf,Gukov:2003na}.
The Chern--Simons function is defined, using a connection (or gauge field) $A$ on an $SU(2)$-bundle $E$ over a three manifold $M$, as
\begin{equation}
    W(A) = \frac{1}{4\pi} \int_M \Tr \left[ A \wedge dA + \frac{2}{3} A \wedge A \wedge A \right] ~.
\end{equation}
The trace is taken in the fundamental representation.
The path integral of Chern--Simons gauge theory is then given in the compact case by
\begin{equation}
    Z(M) = \int_\mathcal{U} [\mathcal{D}A]\ \exp(i k W(A)) ~,
\end{equation}
where $\mathcal{U}$ is the space of $SU(2)$ gauge fields modulo gauge transformations.
The coupling $k$ is integer-quantized, $k \in \mathbb{Z}$, to ensure gauge invariance.
For $SU(2)$ Chern--Simons theory on $M=S^3$, it was shown in~\cite{Witten:1988hf} that the expectation value of a Wilson loop operator along the knot, defined by
\begin{equation}
   U_R(K) = \Tr_R \mathcal{P} \exp \left( - \oint_K A \right)  ~,
\end{equation}
is related to the colored Jones polynomials $J_R(K;q)$ of a knot $K \subset S^3$ evaluated at $q = e^{2\pi i / (k+2)}$.
In our work, we will be interested in evaluations of the Jones polynomial (where the representation $R$ is the fundamental one) away from this particular root of unity, and indeed away from all roots expressible as $e^{2\pi i / (k+2)}$ for some $k \in \mathbb{Z}$.
Strictly speaking, these evaluations of the Jones polynomial are not provided by the usual formulation using the path integral of Chern--Simons theory.
However, evaluation at arbitrary phases $e^{2\pi i /(k+2)}$ for $k \in \mathbb{R}$ can be achieved by making use of the analytic continuation machinery developed in~\cite{Witten:2010cx}.

We will also be interested in a complex-valued Chern--Simons function $W(\mathcal{A})$ obtained from an $SL(2,\mathbb{C})$-bundle $E_\mathbb{C}$ over $M$ with connection $\mathcal{A}$.
In the non-compact $SL(2,\mathbb{C})$ case, there are two couplings, and the path integral is
\begin{equation}
    \mathcal{Z}(M) = \int_{\mathcal{U}_\mathbb{C}} [\mathcal{D}\mathcal{A}][ \mathcal{D} \overline{\mathcal{A}}]\ \exp \left[ \frac{i t}{2} W(\mathcal{A}) + \frac{i \widetilde{t}}{2} W(\overline{\mathcal{A}}) \right] ~,
\end{equation}
where $\mathcal{U}_\mathbb{C}$ is the space of $SL(2,\mathbb{C})$ gauge fields modulo gauge transformations, and the complex couplings $t = \ell + i s$ and $\widetilde{t} = \ell - i s$ obey $s \in \mathbb{C}$ and $\ell \in \mathbb{Z}$.
The coupling $\ell$, which multiplies the real part $\text{Re}(W)$, is integer-quantized for the same reason as $k$ was in the compact case.
On the other hand, $\text{Im}(W)$ is a well-defined complex number even under arbitrary gauge transformations, so $s$ can (in principle) take any complex value.\footnote{Usually one takes $s \in \mathbb{R}$ to ensure that the argument of the path integral is bounded, but via analytic continuation one can obtain something sensible for more general $s$~\cite{Witten:2010cx}.}
There is a particularly interesting saddle point (flat connection) which contributes to $\mathcal{Z}(M)$ in the case where $M$ admits a complete hyperbolic metric.
Such manifolds are relevant for us because, as explained by Thurston~\cite{Thurston1997}, most knots $K$ admit a complete hyperbolic metric on their complements $S^3 \setminus K$.
The knot complement $S^3 \setminus K$ is the three manifold obtained by drilling out a tubular neighborhood around the knot $K \subset S^3$.
This complement is topologically distinct from $S^3$, and the knot shrinks to a torus cusp in the complete hyperbolic metric.\footnote{We emphasize that this complete hyperbolic metric on $S^3 \setminus K$ does not descend somehow from a metric on $S^3$.  It exists due to an ideal tetrahedral decomposition of $S^3 \setminus K$, where each tetrahedron has an embedding in $\mathbb{H}^3$, and the embeddings can be glued together in a consistent way.}
For such hyperbolic $M$, there exists a ``geometric'' flat $SL(2,\mathbb{C})$ connection $\mathcal{A}_-$\footnote{We choose this notation to match (5.61) of~\cite{Witten:2010cx}, and comment further in footnote~\ref{foot:conj}.} for which Im$(W(\mathcal{A}_-))$ is related to the volume of the hyperbolic metric on $M$ via Im$(W(\mathcal{A}_-)) = \text{Vol}(M)/2\pi$.\footnote{The real part Re$(W(\mathcal{A}_-))$ of the geometric connection is proportional to a number which is often called the ``Chern--Simons invariant'' of $M$.}
Thus, $SL(2,\mathbb{C})$ Chern--Simons theory is intimately related to the hyperbolic volumes of three manifolds, as this quantity makes a saddle point contribution to the path integral.

One of the primary motivations for both this work and the work of~\cite{Witten:2010cx} was the so-called volume conjecture~\cite{Kashaev1997}, which relates the colored Jones invariants and the hyperbolic volume of three manifolds (which, by the Mostow--Prasad rigidity theorem, is a topological invariant of any hyperbolic $M$).
As written in the introduction and reproduced here, the volume conjecture states~\cite{Kashaev1997,Murakami2001,Gukov:2003na}
\begin{equation}
    \underset{n \to \infty}{\lim} \frac{2\pi \log |J_n(K;e^{2\pi i /n})|}{n} = \text{Vol}(S^3 \setminus K) ~,
\end{equation}
where $n$ is the $n$-dimensional irreducible representation of $SU(2)$.
Thus, the natural unifying language for the volume conjecture is $SU(2)$ and $SL(2,\mathbb{C})$ Chern--Simons theory, because the knot invariants appearing on both the left and right hand sides are quantities which appear in the calculation of the Chern--Simons path integral.\footnote{This combination of invariants, the colored Jones polynomial and hyperbolic volume, was explored in the context of entanglement entropy of Euclidean path integral states in~\cite{Balasubramanian:2016sro,Balasubramanian:2018por}.}

In writing the above relationship, we must be careful to specify precisely what we mean by the colored Jones invariants $J_n(K;q)$.
We choose a normalization of the colored Jones invariants so that they are Laurent polynomials in the variable $q$, and that the unknot $0_1$ obeys $J_n(0_1;q) = 1$.
As explained clearly in Section 2.5.3 of~\cite{Witten:2010cx}, this choice implies that the colored Jones polynomials are reproduced by the ratio of two Chern--Simons path integrals.
The numerator of this ratio has a single Wilson loop insertion along the knot $K \subset S^3$, and the denominator has a Wilson loop along an unknot $0_1 \subset S^3$.
Explicitly, we have
\begin{equation}
    J_n(K;q=e^{2\pi i /(k+2)}) = \frac{\int_\mathcal{U} [\mathcal{D}A]\  U_n(K) \exp(ikW(A)) }{\int_\mathcal{U} [\mathcal{D}A]\  U_n(0_1) \exp(ikW(A)) } ~,
\label{eq:math-defn}
\end{equation}
and the manifold implicit in $W(A)$ is $M=S^3$.
It is for this reason that the statement of the volume conjecture only involves evaluation of the $n$-colored Jones polynomial at integer $k=n-2$, whereas to understand the conjecture in Chern--Simons theory it is necessary to analytically continue the path integral away from integer values of $k$.
Without normalization, the path integral yields an expression which vanishes at $q=e^{2\pi i /n}$, and this vanishing is removed by dividing by the unknot expectation.
In short, we can either measure how fast the path integral vanishes by computing a derivative with respect to $k$ (which relies on the analytic structure of the function), or we can explicitly divide by a function which vanishes equally as fast to obtain a finite ratio.\footnote{This subtle point, while discussed in this language in~\cite{Dimofte:2010ep}, is implicit throughout~\cite{Witten:2010cx} (also see footnote 7 in~\cite{Dimofte:2010ep}, and note that our $J_n(K;q)$ would be written as $J_n(K;q)/J_n(0_1 ;q)$ in the notation of~\cite{Dimofte:2010ep}).}

\subsection{Analytic continuation}

With these conventions in hand, we will provide a brief review of the main results in~\cite{Witten:2010cx}, essentially to introduce the language, as we will make some speculative comments on the relationship between our numerical results and the analytic techniques in~\cite{Witten:2010cx}.
To analytically continue $SU(2)$ Chern--Simons theory on $S^3\setminus K$ away from integer values of $k$,~\cite{Witten:2010cx} instructs us to first rewrite the $SU(2)$ path integral over $A$ as an $SL(2,\mathbb{C})$ path integral over $\mathcal{A}$ restricted to a real integration cycle $\mathcal{C}_\mathbb{R}$ in the space of complex-valued connections modulo gauge transformations $\mathcal{U}_\mathbb{C}$.
Analytic continuation occurs then by lifting $\mathcal{C}_\mathbb{R}$ to a cycle $\mathcal{C}$ in the universal cover $\widehat{\mathcal{U}}_\mathbb{C}$,\footnote{This space can also be thought of as the space of complex-valued gauge fields modulo topologically trivial, or ``small,'' gauge transformations.} where the Chern--Simons function $W(\mathcal{A})$ is well-defined as a complex-valued function.
(Recall that large gauge transformations can modify the real part of $W(\mathcal{A})$ by a multiple of $2\pi$.)
In the presence of a Wilson loop along a knot, the relevant path integral looks like
\begin{equation}
    \int_{\mathcal{C} \subset \widehat{\mathcal{U}}_\mathbb{C}} [\mathcal{D}\mathcal{A}] \int [\mathcal{D}\rho]\ \exp ( \mathcal{I} ) ~, \qquad \mathcal{I} = ikW(\mathcal{A}) + i I_n(\rho,\mathcal{A}) ~.
\end{equation}
We have added an additional term that depends on a field $\rho$, which is associated with the knot itself.
The introduction of this field along with its action $I_n$ is a way to absorb the Wilson loop $U_n(K)$ into the exponential, and makes use of the Borel--Weil--Bott theorem.
We will not provide a discussion of this point, and simply refer interested readers to~\cite{Witten:2010cx}; we will only refer to the total exponential argument $\mathcal{I}$ from now on.
We will just make one important remark concerning $I_n$: when evaluated on a flat connection, $I_n$ is proportional to $n-1$~\cite{Witten:2010cx}.
Therefore, up to an overall rescaling, $\mathcal{I}$ depends only on the combination \begin{equation} 
\gamma \equiv \frac{n-1}{k} ~,
\end{equation}
when evaluated on a flat connection.\footnote{This differs from the definition of $\gamma$ in~\cite{Witten:2010cx}, where $n/k$ was used instead.  In that discussion, the semiclassical limit was of much greater importance, and in that limit our definition becomes equal to $n/k$.  However, as we are working at $n$ and $k$ of $O(1)$, we will keep the exact ratio implied by the value of $\mathcal{I}$ on a flat connection.}
If we wish to understand the volume conjecture, $\gamma = 1$ is held fixed in the semiclassical limit $n,k\to \infty$.
When quoting values for $\gamma$, we have in mind the ratio~\eqref{eq:math-defn}; in the bare path integral setup of~\cite{Witten:2010cx}, we would need to move slightly away from $\gamma=1$.\footnote{Actually, even with the ratio of path integrals, we need to move away from exactly $\gamma=1$.  We will continue to write $\gamma=1$ as the relevant point for the volume conjecture in the semiclassical limit, but the true value for integer $n$ and $k$ is more like $\gamma = \frac{n-1}{n-2} > 1$.}

The cycle $\mathcal{C}$ must be extended using the machinery of Morse theory; this extension guarantees that the Ward identities will hold.
Morse theory on $\widehat{\mathcal{U}}_\mathbb{C}$, specifically with $\text{Re}(\mathcal{I})$ as a Morse function, plays a key r\^ole in this extension and in the definition of other integration cycles.
Analytic continuation away from a discrete set of points (integer values of $k$, in this case) is not unique, and this corresponds to an ambiguity in lifting $\mathcal{C}_\mathbb{R}$ to $\mathcal{C}$.
The relatively natural resolution in this situation is to ask that the path integral should have no exponentially growing contributions as $k\to \infty$ with fixed $n$.\footnote{As mentioned in~\cite{Witten:2010cx}, it is not quite clear how to enforce this condition on $\mathcal{C}$ for general knots, but we will not need the details.}
This is equivalent to requiring that the colored Jones polynomials, as defined in the mathematical literature, and the ratio of Chern--Simons path integrals~\eqref{eq:math-defn} should hold for more general $q$ after replacing the path integrals with their analytic continuations.\footnote{Conventions are not uniform across the mathematical literature, and the relevance of framing is often unmentioned.  See Appendix~\ref{sec:norm} for further discussion of the alignment between the path integral ratio~\eqref{eq:math-defn} and the mathematical definitions of $J_2$.}

Once the cycle $\mathcal{C}$ has been defined, we must vary the value of $\gamma$ from zero to our desired point.
We begin at zero since we have defined a sort of boundary condition on $\mathcal{C}$ at $k \to \infty$ with fixed $n$ by selecting the relatively natural analytic continuation just described.
As we vary $\gamma$, we must track the behavior of $\mathcal{C}$.
It may seem like there is nothing to keep track of, but in fact there are subtle Stokes phenomena which must be taken into account, as we will now briefly explain.
The cycle $\mathcal{C}$ has a decomposition in terms of so-called Lefschetz thimbles $\mathcal{J}_\sigma$, which are cycles defined using Morse theory that each pass through precisely one critical point $\sigma$ of $\mathcal{I}$, and are defined so that the path integral along them always converges:
\begin{equation}
    \mathcal{C} = \sum_\sigma \mathfrak{n}_\sigma \mathcal{J}_\sigma ~.
\end{equation}
These thimbles can intuitively be visualized as downward flow lines coming from a critical point.
Since our Morse function is $\text{Re}(\mathcal{I})$, the path integral decreases exponentially when following a downward flow line: it is for this reason that convergence is guaranteed.\footnote{The downward flow equations of Morse theory on the space of complex-valued gauge fields, which define the Lefschetz thimbles in the infinite-dimensional path integral setting, behave similarly to the finite-dimensional case due to their elliptic nature.}

When crossing certain loci in the complex $\gamma$ plane, known as Stokes lines, the decomposition of $\mathcal{C}$ in terms of Lefschetz thimbles may be required to change in order to both preserve the cycle $\mathcal{C}$ locally and ensure convergence of the path integral.
Functionally, the coefficients $\mathfrak{n}_\sigma$ change in exactly the right way to compensate for a similar jumping in the Lefschetz thimbles $\mathcal{J}_\sigma$ themselves.
This jumping occurs for a cycle $\mathcal{J}_\sigma$ when there is a downward flow from $\sigma$ that ends at another critical point rather than flowing to $-\infty$.
Thus, recalling that critical points $\sigma$ of $\mathcal{I}$ on $\widehat{\mathcal{U}}_{\mathbb{C}}$ are flat $SL(2,\mathbb{C})$ connections on $S^3 \setminus K$ with a prescribed monodromy around $K$ due to the Wilson loop, Stokes phenomena can lead to the addition of complex $SL(2,\mathbb{C})$ connections to the analytically continued $SU(2)$ path integral, even though we begin with an integration contour that includes only real $SU(2)$-valued connections.

As $\gamma$ is varied, two flat connections can become coincident, which leads to a singularity in the moduli space of flat connections.
Studying such singularities is necessary to understand the Stokes phenomena involved, as there is a trivial solution of the Morse theory flow equations when two flat connections are coincident.
Indeed, the existence of finite-dimensional local models of such singularities allows one to understand some Stokes phenomena in Chern--Simons theory without dealing with the full geometry of $\widehat{\mathcal{U}}_{\mathbb{C}}$.
The point we emphasize here is that, at least for these Stokes phenomena in particular, we may analyze Stokes curves purely as a function of $\gamma$ rather than some more complicated parameter, since the flow is trivial and the only relevant evaluations of $\mathcal{I}$ are on flat connections.

As an explanation for the volume conjecture, we should find that Stokes phenomena in Chern--Simons theory can lead to exponential growth of the path integral.
The final crucial detail which leads to exponentially growing contributions is as follows.
When passing from the space of $SL(2,\mathbb{C})$ gauge fields modulo gauge transformations to its universal cover, each flat connection in $\mathcal{U}_\mathbb{C}$ is lifted to an infinite family of flat connections in $\widehat{\mathcal{U}}_\mathbb{C}$ which differ only in the real parts of their Chern--Simons functions $W$ by a multiple of $2\pi$.\footnote{There is actually a similar issue which arises for the field $\rho$ appearing in the action $I_n$, if we wish to analytically continue in $n$ as well as $k$.  If we do not continue in $n$, this issue modifies the way that downward flow conserves $\text{Im}(\mathcal{I})$.  But again, we refer interested readers to~\cite{Witten:2010cx}.} 
As $\gamma$ is varied, degenerate pairs of these lifted critical points can separate (say at $\gamma_1$), and subsequently recombine (at $\gamma_2$) into new degenerate pairs of two critical points which were not paired initially.

If the Lefschetz thimbles associated to such a pair of critical points are added to the integration cycle due to Stokes phenomena at or before $\gamma_1$, their contributions will have changed drastically by the time $\gamma_2$ is reached.
Namely, the thimbles are now associated with one critical point each from two newly recombined pairs, and this gives in the path integral a difference of phases which vanishes only for $k\in \mathbb{Z}$. 
The prefactor of this phase difference can be exponentially large in $k$, and so the total contribution may diverge exponentially for non-integer $k$.
Indeed, schematically the path integral will have a semiclassical term of the form
\begin{eqnarray}
    e^{ikW(\mathcal{A})}(1-e^{2\pi i k}) ,
\label{eq:semiclassical-form}
\end{eqnarray}
so the pair of critical points will exactly cancel for $k \in \mathbb{Z}$, and diverge exponentially in $k$ for $\text{Im}(W(\mathcal{A})) < 0$ and non-integer real $k>0$.
Therefore, it is the combination of the lifting of critical points to $\widehat{\mathcal{U}}_{\mathbb{C}}$, their splitting and recombination as a function of $\gamma$, and Stokes phenomena which can lead to the situation predicted by the volume conjecture: an $SL(2,\mathbb{C})$ flat connection contributes an exponentially growing term to the asymptotic behavior ($\gamma \to 1$, $k\to \infty$) of an $SU(2)$ Chern--Simons path integral.

We return to these ideas in Section~\ref{sec:cs-implications}, where they become relevant in light of our approximation formula~\eqref{eq:result} and generalizations thereof.

\section{Machine learning}\label{sec:ml}
In this work, we build upon the findings of~\cite{Jejjala:2019kio}, often by employing deep learning~\cite{LeCun2015} (and other machine learning techniques) to decipher the relationships between knot invariants.
Neural networks have been recently employed in knot theory to calculate invariants like the slice genus~\cite{hughes2016neural} and to solve computationally complex problems like unknot recognition~\cite{Gukov:2020qaj}.
Indeed, it is known that a neural network of suitable size can approximate any function~\cite{Cybenko1989,hornik1991approximation}.
Our dataset (which matches that of~\cite{Jejjala:2019kio}) consists of the Jones polynomial, hyperbolic volume, and other knot invariants for all $1,701,913$ hyperbolic knots with $16$ or fewer crossings~\cite{hoste1998first}, tabulated using a combination of the Knot Atlas database~\cite{KnotAtlas} and the \texttt{SnapPy} program~\cite{SnapPy}.

In~\cite{Jejjala:2019kio}, a neural network was used to demonstrate that there is a relationship between the Jones polynomial and the hyperbolic volume of a knot. This was initially achieved with a fully connected two hidden layer network with $100$ neurons in each layer. Experiments with the Jones polynomial evaluations (see Section~\ref{sec:deep}) were initially performed with a network with two hidden layers $50$ neurons wide. Later experiments (see Appendix~\ref{sec:more_exp} for details) found that the network could predict the volumes with roughly $96\%$ accuracy with a two hidden layer network only $5$ neurons wide.
\ref{fig:basic-nn}
The robust performance of these small neural networks is compelling evidence that a simple approximate function exists which maps the Jones polynomial to the hyperbolic volume. Ideally, we would like to go beyond demonstrating the existence of a relationship and actually write down a function that faithfully models the relationship. Unfortunately, the neural network is not much help here. Though it essentially optimizes a function that fits the data, the function is computed via the multiplication and addition of matrices; even in the case of a small network, the multiplication of $5\times5$ matrices still produces functions which are difficult to interpret.

Before describing our strategies for extracting simpler expressions for the relevant correlations learned by the neural networks, we review the deep learning ideas involved.

\subsection{Neural networks}
\begin{figure}
    \centering
    \includegraphics[scale=0.4]{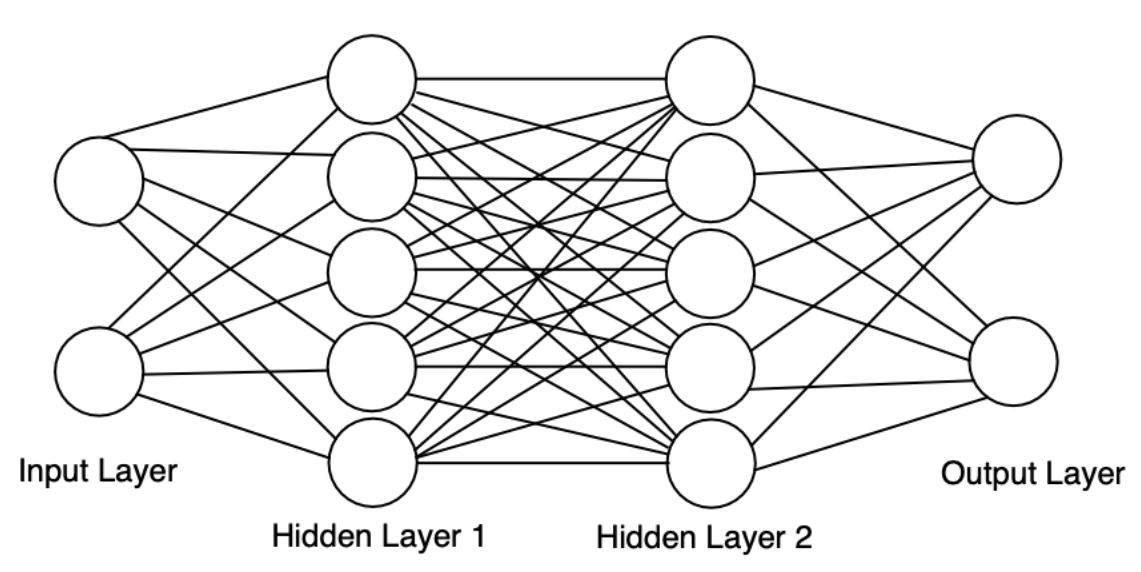}
    \caption{\small{An example of a two hidden layer fully connected neural network architecture. Each hidden layer represents matrix multiplication by a weight vector, the addition of a bias vector, and the element-wise application of an activation function, which introduces non-linearity into the function. In this work we used the Rectified Linear Unit (ReLU) activation function.}}
    \label{fig:neural_net}
\end{figure}

A neural network is a function $f_\theta$ which approximates the relationship between a vector of input features $\mathbf{v}_\text{in}$ and some output $\mathbf{v}_\text{out}$. The network $f_\theta$ is an approximation of the true relationship $A: \mathbf{v}_\text{in} \to \mathbf{v}_\text{out}$. In our case, the input vectors are (vectors based on) the Jones polynomials and the outputs are the corresponding hyperbolic volumes. The dataset is divided into training and testing sets. The network uses the training data to adjust the parameters $\theta$ (the weights and biases) to approximate the relationship $A$ as closely as possible. To do this, a loss function is chosen and minimized in the space of parameters.In the architecture used in this work (see Figure~\ref{fig:neural_net}), the network is built out of $n$ hidden layers which perform matrix multiplication by a weight matrix $W_\theta^m$, addition of a bias vector $\mathbf{b}_\theta^m$, and element-wise application of the activation function $\sigma$. The network can be written as
\begin{equation}
f_\theta(\mathbf{v}_{\text{in}}) = L_{\theta}^n \left( \sigma \left( \ldots L_\theta ^2 \left( \sigma \left( L_\theta^1(\mathbf{v}_{\text{in}}) \right) \right) \ldots \right) \right) ~, \qquad L_\theta^m(\mathbf{v}) = W_\theta^m \cdot \mathbf{v} + \mathbf{b}_\theta ^m ~.    
\end{equation}
The values of hidden layer neurons after applying the activation function are often called the activations $\textbf{a}^m = \sigma(W^m_\theta \cdot \textbf{a}^{m-1} + \textbf{b}^m_\theta)$, with $\textbf{a}^0 \equiv \textbf{v}_{\text{in}}$.
In this work, we use the Rectified Linear Unit (ReLU) activation function, which is $\sigma(x) = x \Theta(x)$, where $\Theta(x)$ is the Heaviside step function.
The loss function is minimized on the training data by using the backpropagation algorithm. This algorithm computes gradients of the loss function for each training data point and adjusts the parameters layer by layer in the network. Once this training is complete, $f_\theta$ is applied to the previously unseen testing set to see how well it approximates $A$. A specific discussion of the neural network architecture used in this work is included in Appendix~\ref{sec:nnsetup}.

\subsection{Layer-wise relevance propagation}
Layer-wise relevance propagation (LRP) is a technique which attempts to explain neural network predictions by calculating a relevance score for each input feature~\cite{bach2015pixel}. This is a useful tool when attempting to derive an analytic function: we can determine which input variables typically carry the most importance when predicting the output. This allows us to hypothesize how to weight input variables in our function and perhaps reduce the complexity of the problem by eliminating redundant variables.

LRP propagates backwards through a neural network, starting at the output layer. The LRP algorithm redistributes the relevance scores from the current layer into the previous layer, employing a conservation property. Denote the activation of a neuron $i$ in layer $m$ by $a_i^m$. Suppose we have all the relevance scores for the current layer, and want to determine the scores $R_j^{m-1}$ in the previous layer. The most basic LRP rule calculates these relevance scores using the formula~\cite{montavon2019layer}
\begin{equation}
    R_j^{m-1} = \sum_k \frac{a_j^{m-1} W_{jk}^m + N_{m-1}^{-1} b_k^m}{\sum_{l}a_l^{m-1} W_{lk}^m + b_k^m}R_k^m ~,
\end{equation}
where the $m$\textsuperscript{th} layer has $N_m$ neurons.
The subscripts on the weights $W$ and biases $b$ here denote matrix and vector indices.
The numerator is the activation of the $j$\textsuperscript{th} neuron in layer $m-1$, multiplied by the weight matrix element connecting that neuron to neuron $k$ in layer $m$ to model how much neuron $j$ contributes to the relevance of neuron $k$.
This fraction is then multiplied by the relevance of neuron $k$ in the layer $m$.
Once the input layer (layer zero) is reached, the propagation is terminated, and the result is a list of relevance scores for the input variables.
The sum in the denominator runs over all of the neurons in layer $m-1$, plus the bias in layer $m$: it imposes the conservation property because we begin with $R^n = 1$ at the output layer and always preserve
\begin{equation}
    \sum_k R_k^m = 1 ~.
\end{equation}
This methodology was originally proposed in a classifier problem; we have adapted it to the case of regression.

\subsection{Strategies for analyzing the network}

Armed with the evidence that our function exists, how do we proceed? The input is too complicated for educated guesswork or traditional curve fitting techniques, as our encoding of the Jones polynomial is a $16$-vector.\footnote{In~\cite{Jejjala:2019kio}, we provided the degrees of the Jones polynomial as inputs. As we explicate in Appendix~\ref{sec:repofJ}, the neural network performs just as well with the coefficients alone. This is the $16$-vector to which we refer.} Our approach involves performing experiments through which we can probe how the neural network makes its predictions. As stated previously, a neural network's inner workings are largely inaccessible. Despite this, by studying a network's success when faced with a transformed or truncated form of the Jones polynomial, we can begin to update our idea of what the eventual function might look like. There are three main ways that we accomplish this.

The first type of experiment is training a neural network on some truncated or mutated form of the input data. For instance, inspired by the volume-ish theorem~\cite{Dasbach_2007}, we could create a new input vector containing just the degrees of the polynomial and the first and last coefficients: $\left( p_{\text{min}},  p_{\text{max}}, c_1, c_{-1} \right)$. If the neural network was then still able to learn the volume, with comparable accuracy to the original experiment, then perhaps our function could be built from these four numbers. 
It turns out that the neural network did not perform well with this input. Another example of this method is detailed in Section~\ref{sec:deep}. 

The second strategy is taking a neural network which has already been trained and feeding it altered input data. For instance, if we train a network on our original $16$-vectors and give it an input vector where certain elements have been zeroed out or shifted by some constant, can it still predict the volume? This allows us to probe our function for, among other things, redundant variables and invariance under translations.
We could, for example, shift pairs of input variables by a constant and record the network's ability to predict the volume. These experiments are inspired by~\cite{Udrescu:2019mnk}, where symbolic regression and machine learning are approached with a physicist's methodology.\footnote{We explore symbolic regression techniques further in Appendix~\ref{sec:symbolic}.  We obtain $96.56$\% accuracy, but the formul\ae\ are not obviously interpretable.}

Together with these two experiments, we use LRP to understand the relevance of the various input features in making a prediction. 
As we reviewed above, LRP is an algorithm which uses the neuron activations, along with the weight and bias matrices, to map the activations in the final network layer back onto the input layer. This technique is successfully implemented in Section~\ref{sec:deep} to further reduce the number of input variables needed to predict the volume, eventually yielding formulas with just a handful of numerical parameters and a single nonlinearity.

\section{An approximation formula for the hyperbolic volume}\label{sec:deep}

Our goal is to employ the machine learning techniques reviewed in Section~\ref{sec:ml} to determine what particular property of the Jones polynomial was exploited by the neural network in~\cite{Jejjala:2019kio} to compute the hyperbolic volume.
Inspired by the observations of~\cite{Witten:2010cx} concerning the analytic continuation of Chern--Simons theory that we briefly summarized in Section~\ref{sec:knots}, we approach this task by evaluating the Jones polynomial at various roots of unity.
Indeed, the Jones polynomial is determined by its values at roots of unity, so we lose no essential information if we include enough such evaluations.
We use techniques from interpretable deep learning to reverse engineer neural networks in Section~\ref{sec:interpretable} and comment on the implications for the path integral of analytically continued Chern--Simons theory in Section~\ref{sec:cs-implications}.

\subsection{Interpretable deep learning}\label{sec:interpretable}

We begin by training a neural network composed of two fully connected hidden layers with $50$ neurons each on the Jones polynomial evaluated at $e^{2\pi i p /(r+2)}$, for integers $r \in [3, 20]$ and $p \in [0, r+2]$.
Complex conjugates are omitted from this tabulation since Laurent polynomials obey $J_2(K;\overline{q}) = \overline{J_2(K;q)}$.
The input data (which includes all hyperbolic knots up to and including $15$ crossings) is represented as a vector where the entries at position $2p$ and $2p+1$ correspond to the real and complex parts of the $p$\textsuperscript{th} evaluation.  
Layer-wise relevance propagation (LRP)~\cite{bach2015pixel, montavon2019layer} is used to identify which evaluations of the Jones polynomial are important in calculating the volume.\footnote{Layer-wise relevance propagation has not been widely applied in the physics context, but see~\cite{Bluecher:2020kxq} for an example.}
 
\begin{figure}[t]
   \centering
\begin{tabular}{ccccc}
\includegraphics[scale=0.25]{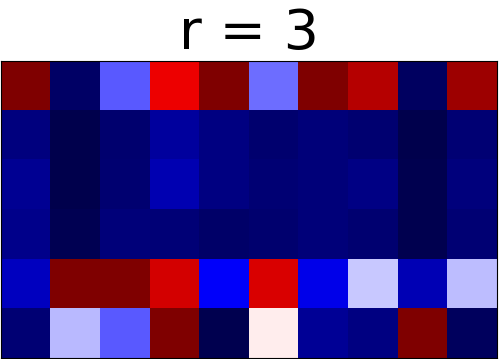}&
\includegraphics[scale=0.25]{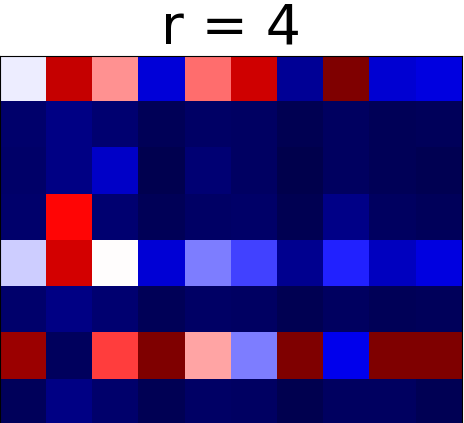}&
\includegraphics[scale=0.25]{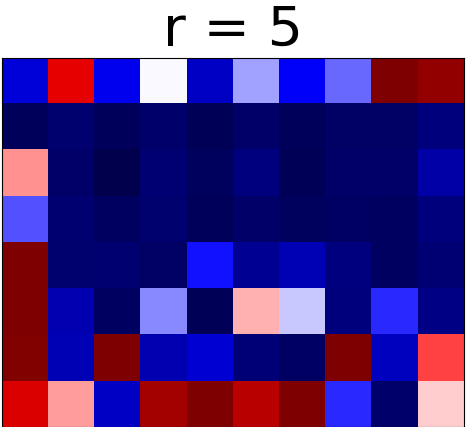}&
\includegraphics[scale=0.25]{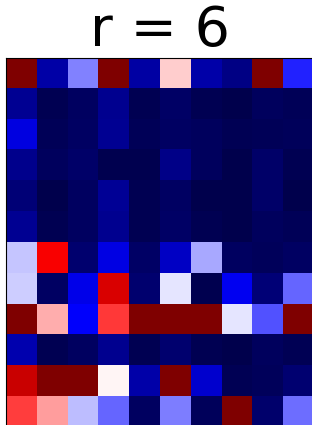}&
\includegraphics[scale=0.25]{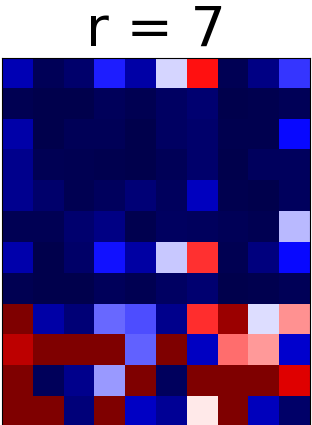}
\end{tabular}
    \caption{\small{Each grid shows the LRP results for $10$ neural network inputs. The grid labels $r \in [3,7]$ determine the form of the input vector. Each column represents a single input, corresponding to the real and imaginary parts of the knot's Jones polynomial evaluated at the phases $e^{2\pi i p/(r+2)}$ with $p\in\mathbb{Z}$ and $0\leq 2p \leq r+2$. Dark blue squares correspond to the smallest magnitude for the relevance score, and red to the highest. So, in a given column, the red squares represent evaluations which were highly relevant for the neural network to predict that knot's volume.
    We also see that the same evaluations are often relevant across all $10$ knots in the displayed set; these correspond to mostly red rows. 
    }}
\label{fig:LRP}
\end{figure}

The LRP results are easily interpreted graphically, as demonstrated in Figure~\ref{fig:LRP}. 
We can immediately see which columns are relevant in making a prediction of the volume. 
The final column in Table~\ref{tab:root_table} shows the results when training the neural network on just those evaluations which LRP determines are relevant. Comparing the values in the final column to the original error in the second column, we see that there is no real reduction in performance. 
LRP successfully selects the relevant input features, and in some cases performance is actually improved by pruning irrelevant inputs.

\begin{table}[t]
\centering
    \begin{tabular}{|c | c | c |c|c|}
    
        \hline
        $r$ & Error & Relevant roots& Fractional levels & Error (relevant roots)\\
        \hline
        $3$ & $3.48\%$ & $e^{4\pi i/5}$ & $\frac{1}{2}$& $3.8\%$\\
        \hline
        $4$ & $6.66\%$ & $-1$ & $0$& $6.78\%$\\
        \hline
        $5$ & $3.48\%$ & $e^{6\pi i/7}$ & $\frac{1}{3}$& $3.38\%$\\
        \hline
        $6$ & $2.94\%$ & $e^{3\pi i/4}$, $-1$ & $\frac{2}{3}$, $0$& $3\%$\\
        \hline
        $7$ & $5.37\%$ &  $e^{8\pi i/9}$ & $\frac{1}{4}$& $5.32\%$\\
        \hline
        $8$ & $2.50\%$ & $e^{3\pi i/5}$, $e^{4 \pi i /5}$, $-1$ & $\frac{4}{3}$, $\frac{1}{2}$, $0$& $2.5\%$\\
        \hline
        $9$ & $2.74\%$ & $e^{8\pi i/11}$, $e^{10\pi i/11}$ & $\frac{3}{4}$, $\frac{1}{5}$& $2.85\%$\\
        \hline
        $10$ & $3.51\%$ & $e^{2\pi i/3}$,$e^{5\pi i/6}$, $-1$ & $1$, $\frac{2}{5}$, $0$& $4.39\%$\\
        \hline
        $11$ & $2.51\%$ & $e^{8\pi i/13}$, $e^{10 \pi i/13}$, $e^{12\pi i/13}$ & $\frac{5}{4}$, $\frac{3}{5}$, $\frac{1}{6}$& $2.44\%$\\
        \hline
        $12$ & $2.39\%$ & $e^{5\pi i/7}$, $e^{6\pi i/7}$, $-1$ & $\frac{4}{5}$, $\frac{1}{3}$, $0$& $2.75\%$\\
        \hline
        $13$ & $2.52\%$ &$e^{2\pi i/3}$, $e^{4\pi i/5}$, $e^{14 \pi i/15}$ & $1$, $\frac{1}{2}$, $\frac{1}{7}$& $2.43\%$\\
        \hline
        $14$ & $2.58\%$ & $e^{3\pi i/4}$, $e^{7 \pi i /8}$, $-1$ & $\frac{2}{3}$, $\frac{2}{7}$, $0$& $2.55\%$\\
        \hline
        $15$ & $2.38\%$ & $e^{12\pi i/17}$, $e^{14\pi i/17}$, $e^{16\pi i/17}$&  $\frac{5}{6}$, $\frac{3}{7}$, $\frac{1}{8}$& $2.4\%$\\
        \hline
        $16$ & $2.57\%$ & $e^{2\pi i/3}$, $e^{7 \pi i /9}$, $e^{8\pi i /9}$, $-1$ & $1$, $\frac{4}{7}$, $\frac{1}{4}$,  $0$& $2.45\%$\\
        \hline
        $17$ & $2.65\%$ & $e^{14\pi i/19}$, $e^{16\pi i/19}$, $e^{18\pi i/19}$, & $\frac{5}{7}$, $\frac{3}{8}$, $\frac{1}{9}$& $2.46\%$\\
        \hline
        $18$ & $2.49\%$ & $e^{4\pi i/5}$, $e^{9\pi i/10}$, $-1$ & $\frac{1}{2}$, $\frac{2}{9}$, $0$& $2.52\%$\\
        \hline
        $19$ & $2.45\%$ & $e^{2\pi i/3}$, $e^{16\pi i/21}$, $e^{6\pi i/7}$, $e^{20\pi i/21}$ & $1$, $\frac{5}{8}$, $\frac{1}{3}$, $\frac{1}{10}$& $2.43\%$\\
        \hline
        $20$ & $2.79\%$ & $e^{8\pi i/11}$, $e^{9\pi i/11}$, $e^{10\pi i/11}$, $-1$ & $\frac{3}{4}$, $\frac{4}{9}$, $\frac{1}{5}$, $0$& $2.4\%$\\
        \hline
    \end{tabular}
\caption{\small{LRP results when training the neural network on different roots of unity, corresponding to different $r$. The third column shows the roots relevant to calculating the output, the fourth column shows the corresponding fractional levels $k$ obtained by writing the relevant roots as $e^{2\pi i / (k+2)}$, and the final column the error when training the network on only the relevant roots.
}}
\label{tab:root_table}
\end{table}

The results in Table~\ref{tab:root_table} were each obtained over one run, so fluctuations in the error are expected. 
To test the stability of the results, we trained the neural network $20$ times, using $J_2(K;e^{4\pi i/5})$ (the relevant root for $r=3$) as input. 
Averaged over $20$ runs, the error was $3.71 \pm 0.06 \%$. 
With this relatively small standard deviation, we conclude that network performance is stable over multiple runs.

\begin{figure}[t]
    \centering
    \includegraphics[scale=0.47]{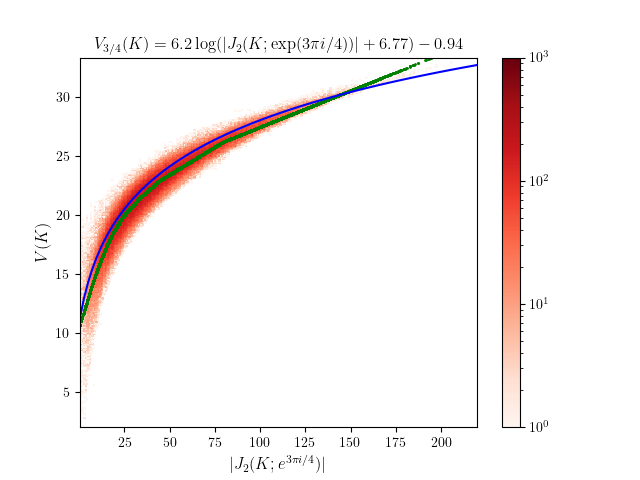}
    \includegraphics[scale=0.47]{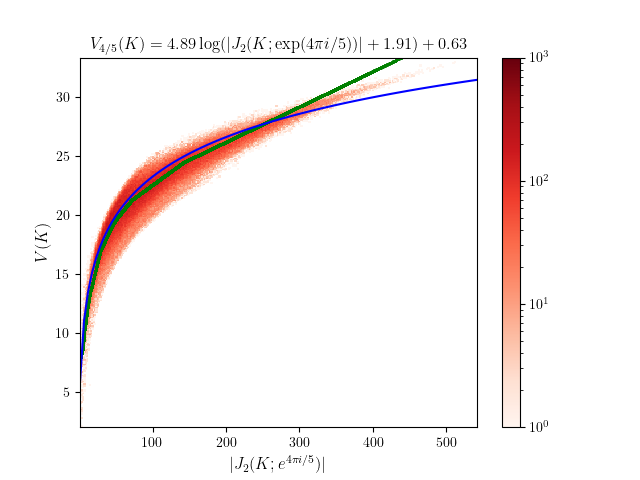}
    \caption{\small{(Left) Best fit for the hyperbolic volume as a function of $|J_2(K;e^{3\pi i/4})|$. The fitted function (blue) predicts the volume to $97.14\%$ accuracy and the neural network trained on the magnitude of $J_2(K;e^{3\pi i/4})$ (green) achieves $96.55$\% accuracy. (Right) Best fit for the hyperbolic volume as a function of $|J_2(K; e^{4\pi i/5})|$. The fitted function (blue) predicts the volume to $96.67\%$ accuracy and the neural network trained on the magnitude of $J_2(K; e^{4\pi i/5})$ (green) achieves $96.30$\% accuracy.
    Both plots display $(J_2,\text{Vol})$ data pairs as a density cloud where darker red regions contain more points.
    }} \label{fig:best_fit_new}
\end{figure}

In the first line of Table~\ref{tab:root_table}, the neural network is learning to predict the volume from two numbers: the real and imaginary parts of $J_2(K;e^{4\pi i/5})$. 
If we represent the input as a magnitude and a phase rather than as a real and imaginary part, the performance of the neural network is unchanged. 
In fact, it turns out that if we drop the phase and work only with the magnitude, the performance remains unchanged; the phase does not matter. 
This means that the network is predicting the volume (to $96.29 \pm 0.06 \%$ accuracy) with just one number: $|J_2(K;e^{4\pi i/5})|$.
Another promising candidate is $|J_2(K;e^{3\pi i /4})|$, which was also determined by relevance propagation to be important in several rows of Table~\ref{tab:root_table}.
Plotting either $|J_2(K;e^{4\pi i/5})|$ or $|J_2(K;e^{3\pi i/4})|$ against the volume (Figure~\ref{fig:best_fit_new}), we find a graph which looks like $a \log(x + b) + c$ for some constant $O(1)$ parameters $a$, $b>0$, and $c$. 
Thus, we may be able to obtain a reasonably compact analytic expression by performing simple regression on such an ansatz.

\begin{figure}[t]
    \centering
    \includegraphics[scale=0.45]{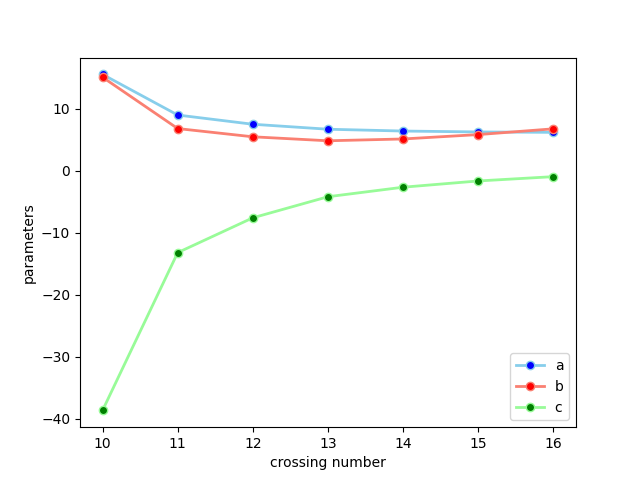}
    \includegraphics[scale=0.45]{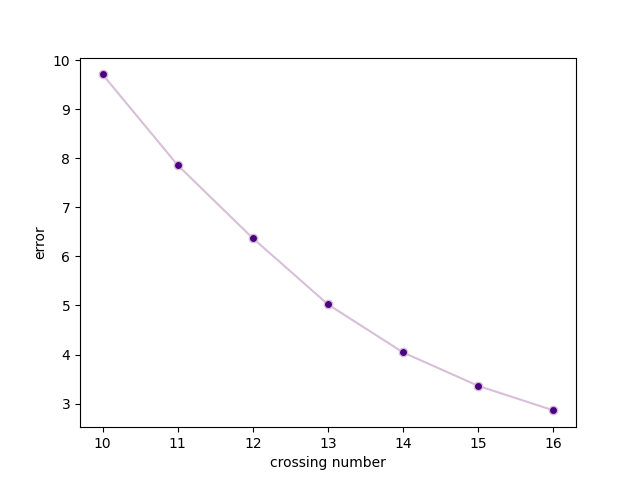}
    \caption{\small{Variation of fit parameters with regression data. As more regression data is included, the fit parameters converge to some constant values. This indicates that the approximation formulas should perform well outside of our dataset.}}
    \label{fig:parameter-variation}
\end{figure}

We use the curve fitting and optimization procedures built into \texttt{Mathematica}. 
The dataset now includes all hyperbolic knots up to and including $16$ crossings. 
For $|J_2(K;e^{3\pi i/4})|$, we find the formula
\begin{equation} 
V_{3/4}(K) = 6.20\log{(|J_2(K;e^{3\pi i/4})| + 6.77)} - 0.94 ~,
\label{eq:3pi4formula}
\end{equation}
which predicts the volume with an error of $2.86\%$ (Figure~\ref{fig:best_fit_new}, left).
This result is comparable to some of the best performances in Table~\ref{tab:root_table} and performs only slightly worse than the neural network which had access to the full Jones polynomial.
Repeating the analysis for $|J_2(K;e^{4\pi i/5})|$, the optimized formula is
\begin{equation} 
V_{4/5}(K) = 4.89\log{(|J_2(K;e^{4\pi i/5})| + 1.91)} + 0.63 ~,
\label{eq:4pi5formula}
\end{equation}
which predicts the volume with an error of $3.33\%$ (Figure~\ref{fig:best_fit_new}, right).
Due to the fact that the error increase is small and the parameter reduction is enormous in passing from the black box neural network to the formulas~\eqref{eq:3pi4formula} and~\eqref{eq:4pi5formula}, we may conclude that we have more or less successfully reverse engineered the simple function which is learned by the network.
The dependence of our approximation formulas only on the absolute value of a Jones polynomial evaluation aligns nicely with the unchanged performance of the neural network when we drop the maximum and minimum degree information or cyclically permute the polynomial coefficients, as both operations leave the absolute value unchanged (see Appendix~\ref{sec:repofJ} for other experiments with input representation).
Moreover, when performing regression on only a subset of the dataset and increasing the size of that subset, the determined coefficients $a$, $b$, and $c$ are roughly converging to some constant values (Figure~\ref{fig:parameter-variation}, left), which gives us confidence that these approximation formulas should perform well even outside of our dataset.
As more knots are included in the regression, the error on the total dataset reaches its optimal value smoothly (Figure~\ref{fig:parameter-variation}, right).
For comparison, a neural network trained on $|J_2(K;e^{3\pi i/4})|$ predicts the volumes to $96.55$\% accuracy, and a neural network trained on $|J_2(K;e^{4\pi i/5})|$ predicts the volumes to $96.30$\% accuracy. 
We observe that at large magnitudes for $J_2(K;e^{i\theta})$, both the best fit function and the neural network start to differ from the true volumes.
This effect is attributable to the sparsity of data.

\begin{figure}[t]
    \centering
    \includegraphics[scale=0.47]{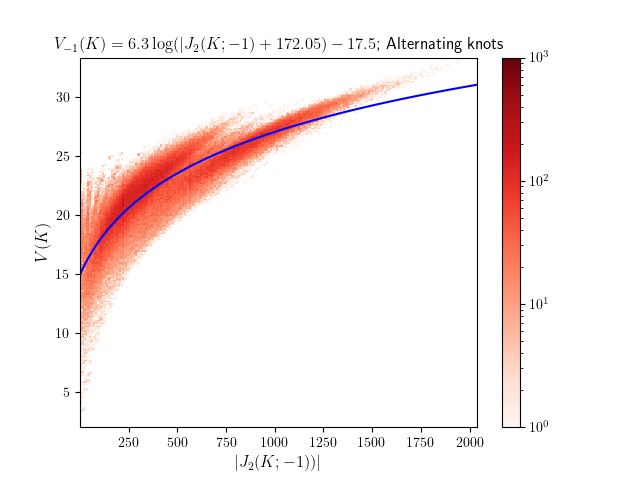}
    \includegraphics[scale=0.47]{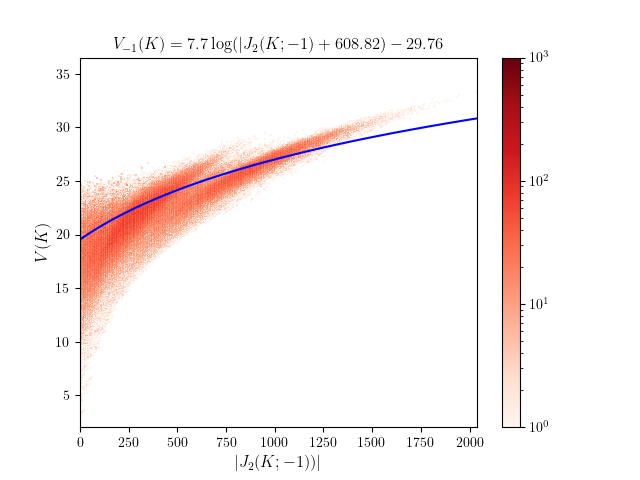}
  \caption{\small{Comparison of~\eqref{eq:3pi4formula} with a fit of the same functional ansatz on $J_2(K;-1)$; as in~\cite{Dunfield2000}. When applied to alternating knots only (left), the function predicts the volume with a mean error of $5.76\%$. Fitting the same ansatz to all knots (right) predicts the volume with an error of $6.16\%$. }}
  \label{fig:Dun_compare}
\end{figure}

In~\cite{Dunfield2000}, it was noted that for alternating knots up to $13$ crossings that $V(K)$ was roughly proportional to $\log |J_2(K; -1)|$.
As a comparison to our results, we fit the functional ansatz $a \log(|J_2(K;-1)| +b) +c$ for alternating and non-alternating knots (Figure~\ref{fig:Dun_compare}). 
For knots up to and including $16$ crossings, the functions predict the volumes of alternating knots to $94.24\%$ accuracy and the volumes of all knots to $93.84\%$ accuracy.
While this phase performs worse than the ones found by our layer-wise relevance propagation, it is worthwhile to understand the extent to which we can vary the particular phase used in this approximation ansatz and still retain reasonable performance, since we did find that other roots of unity are relevant for different values of $r$ in Table~\ref{tab:root_table}. 
\begin{figure}[t]
    \centering
    \includegraphics[scale=0.7]{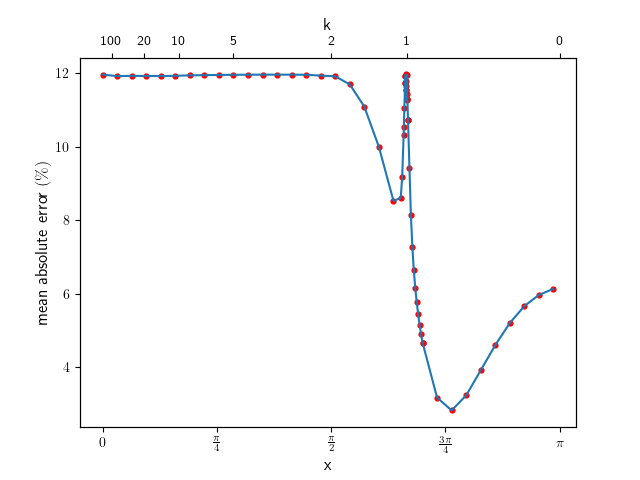}
    \caption{\small{Mean absolute errors as a function of $x$ when optimizing the ansatz $V(K) = a \log(|J_2(K; e^{ix})| + b)+c$.  We re-optimize the constants $a$, $b$, and $c$ for each individual value of $x$.  The minimum error of $2.83\%$ occurs at $x=2.3$. This is evidence that we can approximate the hyperbolic volume from the Jones polynomial evaluated at a wider variety of fractional levels than just $k=\frac{1}{2}$ or $k=\frac{2}{3}$.
    }}
    \label{fig:min_err}
\end{figure}

To explore this possibility, we optimize the ansatz $V(K) = a \log({|J_2(K;e^{ix})|+b)} + c$ in \texttt{Mathematica} and report the mean absolute error under the constraint $x \in [0, \pi]$ (Figure~\ref{fig:min_err}) for all knots up to and including $16$ crossings.\footnote{Amusingly, as is now ubiquitous in high energy theory, the graph evinces the customary dip, ramp, plateau shape~\cite{Cotler:2016fpe}.  Of course, here these features are not connected to random matrix theory in any clear way. A tentative explanation for the location of the dip makes use of the saddle point form of the path integral~\eqref{eq:semiclassical-form}, where there is a reasonable spread of this value for different volumes in the region $k \in [0.5,0.8]$.  However, we are far from the semiclassical limit, and cannot really trust~\eqref{eq:semiclassical-form}. }
The minimum error of $2.83\%$ is realized for $x = 2.3$. We can relate this to the relevant roots in Table~\ref{tab:root_table} by noting that this root is similar to $e^{3\pi i/4}$. This corresponds to a fractional level of $\frac{2}{3}$ appearing in the $r=6$ and $r=14$ lines of Table~\ref{tab:root_table}.
Notice also that there is a plateau in error rate at around 12\% for $x \lesssim \frac{\pi}{2}$.
The error of $12$\% is significant;
if we take $V(K) = V_0$, where $V_0$ is the average volume of all knots  up to $16$ crossings, the error for this constant approximation function is $11.97$\%.
This represents a latent correlation in the dataset, and if we have an error rate greater than or equal to this then our approximation formula is learning essentially nothing.
This error plateau, and other interesting features of Figure~\ref{fig:min_err} like the error spike around $k=1$, will be analyzed in the context of Chern--Simons theory in Section~\ref{sec:cs-implications}.
Before moving on, we note that upon closer inspection of the region around $k=1$, the single error spike shown in Figure~\ref{fig:min_err} is actually resolved into two separate spikes at $k=1$ and $k=1.014$ with a shallow dip of roughly 0.7\%.
We do not have a good understanding of this feature, and we have not ruled out numerical errors as its source, because our \texttt{Mathematica} fitting procedures give convergence warnings at these levels.
The robust conclusion that we claim here is only that there is a large error spike around $k=1$ which reaches up to the height of the plateau.

For completeness, we also record the bounding curves of the dataset, computed using our approximation ansatz.
The curves
\begin{align}
    f(|J_2(K;e^{3\pi i/4})|) &=  6.053 \log(33.883 + |J_2(K; e^{3\pi i/4})|) + 0.798 ~,\\
    g(|J_2(K;e^{3\pi i/4})|) &=  8.210 \log(0.695 + |J_2(K; e^{3\pi i/4})|) -12.240 ~,
\end{align}
completely enclose the hyperbolic volumes of the knots up to and including $16$ crossings (Figure~\hyperlink{4_1}{\includegraphics[scale=.007]{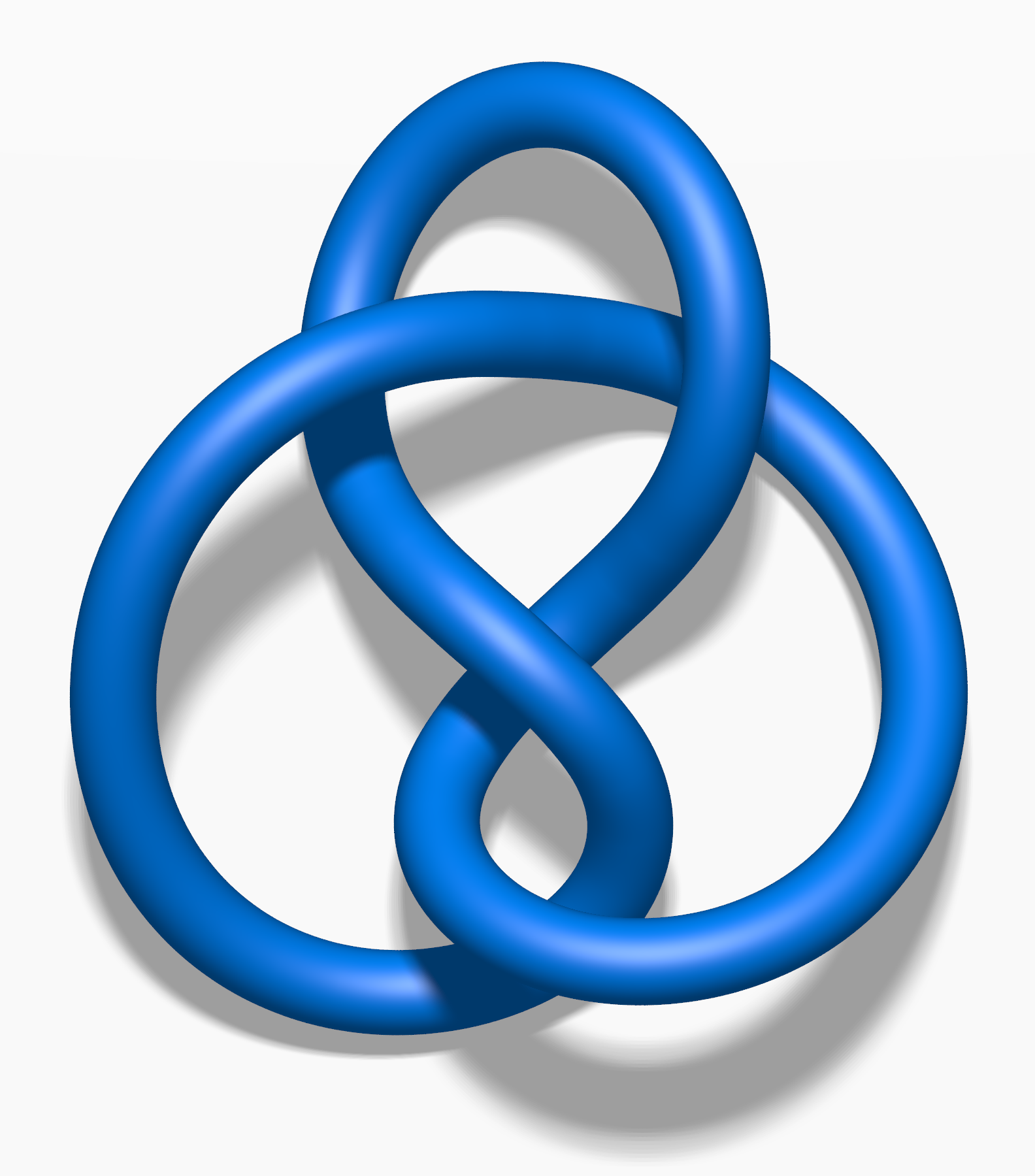}}.) 

\begin{figure}\hypertarget{4_1}
    \centering
    \includegraphics[scale=0.7]{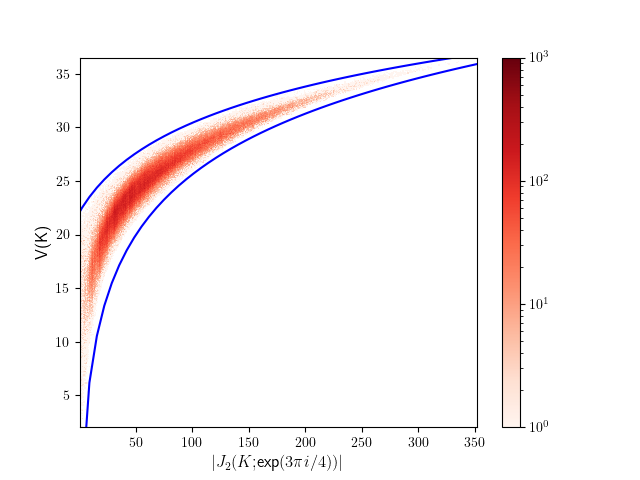}
    \caption*{
    \textbf{Figure} \includegraphics[scale=.025]{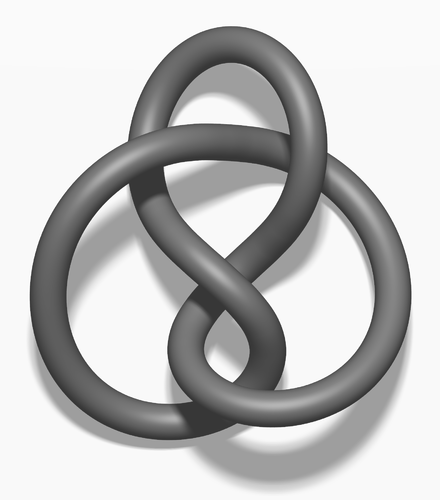}\textbf{:}
    \small{Bounding curves for the hyperbolic volumes up to and including $16$ crossings. }}
    \label{fig:bounds}
\end{figure}\addtocounter{figure}{1}
    
\subsection{Implications in Chern--Simons theory}\label{sec:cs-implications}

Using machine learning and data analysis techniques, we determined that the relevant information about Vol$(S^3\setminus K)$ contained in $J_2(K;q)$ can be efficiently extracted with a single evaluation of $J_2$ at a root of unity $q = e^{2\pi i /(k+2)}$, which corresponds to fractional level $k=\frac{1}{2}$ or $k=\frac{2}{3}$.
At $k=\frac{2}{3}$, for example, we have $q = e^{3\pi i / 4}$, and we find that the simple function~\eqref{eq:result} reproduces the hyperbolic volume of $K$ with almost no loss of accuracy compared to the full neural network of~\cite{Jejjala:2019kio}.
The fractional level $k=\frac{1}{2}$ has been studied in the context of Abelian Chern--Simons theory~\cite{moore2019introduction}, but has not played a significant r\^ole in the non-Abelian theory we study here, and level $k=\frac{2}{3}$ (to our knowledge) has hardly been studied at all.
Of course, due to gauge invariance, non-Abelian Chern--Simons theory cannot be defined in the na\"{\i}ve way at fractional level, and it is for this reason that the analytic continuation techniques of~\cite{Witten:2010cx} have an intriguing bearing on our observations.

Our numerical results suggest the Lefschetz thimbles involved in the analytically continued path integral at $\gamma = 2$ or $\gamma=\frac{3}{2}$ are somehow related to the set of thimbles which appear semiclassically at $\gamma = 1$.
Perhaps the set of relevant thimbles around $\gamma \in (\frac{3}{2},2)$ contains the geometric conjugate $SL(2,\mathbb{C})$ connection that we expect semiclassically at $\gamma = 1$.\footnote{\label{foot:conj}The geometric conjugate connection is a flat $SL(2,\mathbb{C})$ connection which has Im$(W) = -\text{Vol}(M)/2\pi$, which is necessary for the factors of $i$ to cancel and yield $e^{k \text{Vol}(M)/2\pi}$ in a saddle point approximation of $|\mathcal{Z}(M)|$ or the analytic continuation of $|Z(M)|$.  It is simply the complex conjugate of the geometric connection we mentioned previously. To prevent any ambiguity, we remark that the object called $\mathcal{A}_+$ in (5.61) of~\cite{Witten:2010cx} is an instance of what we are calling the geometric conjugate connection.}
This interpretation suggests that the geometric conjugate connection contributes to the path integral for the maximum number of knots in our dataset around $\gamma = \frac{3}{2}$, or $k = \frac{2}{3}$, since that is where the approximation formula performs optimally.
This location coincides with the bottom of the dip in Figure~\ref{fig:min_err}.
Similarly, the ramp between $\frac{2}{3}<k<2$ (ignoring the spike at $k=1$) is consistent with the interpretation that large fractions of knots are beginning to lose the geometric conjugate connection for $k > \frac{2}{3}$.
That the dip begins at $k = 0$ with an error rate which is already fairly low may signal that many knots retain the geometric conjugate connection even as $\gamma$ becomes large.

We must emphasize that we are not working in the semiclassical limit of large $n$ and $k$.
Instead, we are actually in the opposite limit where both quantities are $O(1)$.
Therefore, quantum effects in the path integral are strong, and it is not clear how to isolate the contribution of the critical points.
Nevertheless, it seems that our machine learning techniques suggest this is approximately possible with some simple processing of the path integral result.
This phenomenon is a bit reminiscent of the ease with which one-loop quantum corrections are incorporated in Chern--Simons theory, through a simple shift $k \to k+2$ and $n \to n+1$ in the semiclassical formulas.
Here, we have an effect at the level of the analytically continued path integral, where a simple multiplication and shift can absorb quantum corrections to a reasonable degree.
It would be interesting to try to make this precise.

It is also instructive to study the error spike at $k=1$ in Figure~\ref{fig:min_err}.
Readers familiar with the main message of~\cite{Witten:2010cx} will understand why this feature is in fact expected: at integer values of the Chern--Simons level with $k +1 \geq n $, the path integral ratio~\eqref{eq:math-defn} receives contributions only from $SU(2)$-valued critical points, and can be analyzed using the original techniques in~\cite{Witten:1988hf}.
In other words, the $SU(2)_k$ current algebra has an integrable representation of dimension $n=k+1$, and so the techniques of~\cite{Witten:1988hf} are sufficient and no analytic continuation is necessary.
Of course, at $n = k+2$, there is a vanishing of the bare path integral, but the ratio~\eqref{eq:math-defn} remains finite and is sensitive to the volume like the analytically continued bare path integral.
That this occurs only at the special value $n = k+2$ and not for any $n < k+2$ essentially follows from the fact that the bare expectation value of the unknot does not vanish for $0 < n < k+2$ with integer $k$.

This observation supports our interpretation that the presence of the geometric conjugate connection is responsible for the approximation formula's success. 
In this case, we cannot expect to learn anything about the hyperbolic volume at $k=1$ and $n=2$ beyond latent correlations in the dataset because the geometric conjugate connection does not contribute to the path integral there. 
Even if it is present in the numerator of~\eqref{eq:math-defn}, it will cancel with another contribution for integer $k > n-2$, and so our approximation formula performs as poorly as possible at $k=1$.
However, there are sharp improvements to the approximation formula just below or above this value, where $k$ again becomes fractional and we may find a contribution from the geometric conjugate connection.

Intriguingly, there seems to be a transition around $\gamma = \frac{2}{3}$ (equivalently, $k = \frac{3}{2}$) where the approximation formulas begin to improve over the maximum error, and~\cite{Witten:2010cx} found that the geometric conjugate connection for the figure-eight knot is added to the path integral by a Stokes phenomenon precisely at this value.
The appearance of this maximum error plateau, which roughly matches the error of taking a constant approximation function equal to the average volume in the dataset, is consistent with the interpretation that the approximation formula with $k>\frac{3}{2}$ fails to convey anything useful because the geometric conjugate connection is completely absent from the numerator of~\eqref{eq:math-defn}, rather than being present and canceling for integer $k$.
Perhaps there is a larger class of knots with a Stokes phenomenon at $\gamma = \frac{2}{3}$ which adds the geometric conjugate connection to the path integral, and there are only a few knots (or perhaps none at all) which receive such a contribution for $k > \frac{3}{2}$.

If we follow the suggestion that the success of our approximation formula is signalling the presence of the geometric conjugate connection for most knots in the dataset, we are led to an interesting observation concerning the volume conjecture.
As discussed previously, it is well known that the volume conjecture does not converge monotonically in $n$~\cite{Garoufalidis_2005}.
Perhaps the reason for this is that the geometric conjugate connection is missing for early terms in the sequence, and appears at a certain point, after which convergence is monotonic.
For example, the first term in the sequence involves $|J_2(K;-1)|$, which corresponds to $k=0$ and therefore $\gamma = \infty$.
While it may be that some knots acquire the geometric conjugate contribution by $\gamma = 1$ and never lose it for any $\gamma >1$, this may not be the case for all knots.
If the geometric conjugate connection is lost by Stokes phenomena at some $\gamma > 1$, the first term in the volume conjecture sequence cannot be expected to be a good approximation.
We comment further on this in Section~\ref{sec:disc}.

\section{Discussion}\label{sec:disc}

In this work, we have utilized neural networks and layer-wise relevance propagation to extract a very simple function $V(K)$ which predicts with better than $97\%$ accuracy the volume of a hyperbolic knot using only a single evaluation of the Jones polynomial at a root of unity.
The existence of such a function was predicted in~\cite{Jejjala:2019kio}, and prior related observations had also been suggestive~\cite{Dunfield2000,khovanov2003}.
The main problem described at the end of~\cite{Jejjala:2019kio} was to determine what simple function was being computed by a neural network, and the roughly equal accuracy of our simple function and the neural network is strong evidence that we have succeeded.
We also found excellent alignment between the form of the approximation ansatz and the operations on the inputs (dropping degree information, cyclic coefficient permutations) which left the network performance unchanged.

We commented briefly on the implications of this result for analytically continued Chern--Simons theory in Section~\ref{sec:cs-implications}.
It is clear that there is at least some connection, because the root of unity involved in the calculation is not accessible by the standard definition of the path integral: it corresponds to fractional level $k$.
However, precisely what is gained by studying the path integral far from the semiclassical limit around $\gamma = \frac{3}{2}$ or $\gamma = 2$, and why this should be related to the semiclassical limit near $\gamma = 1$, is not at all clear.
Our usual intuition from quantum field theory suggests that the strong quantum effects at small $k$ ought to completely mask the particular value of the action at the geometric conjugate connection. 
Mysteriously, this does not happen, and (at least, \textit{e.g.}, for large $|J_2(K;e^{3\pi i / 4})|$) there is an essentially universal way to extract the saddle point value of this particular connection with high accuracy for any hyperbolic knot.
We have found some supporting evidence for our interpretation that the success of the approximation formula signals the presence of the geometric conjugate connection in most knots in our dataset.
This evidence involved the spike in error near the integer $k=1$ as well as the rough matching between the location of the relevant Stokes phenomenon for the figure-eight knot and the critical value of $\gamma$ where the approximation formula begins to perform well.

An interesting future direction would be to try to derive an inequality, along the lines of the volume-ish theorem~\cite{Dasbach_2007}, using analytically continued Chern--Simons theory.
Indeed, the volume-ish theorem should generalize to higher colors, with the upper and lower bounds converging in the infinite $n$-limit.
Deducing an inequality seems quite difficult, as the analysis is very involved for each individual knot~\cite{Witten:2010cx}.
Nevertheless, we may hope to find some unifying theme now that we have a specific location of interest (further along the real $\gamma$ axis than previously suspected).
As a very first step, one would have to understand how to evaluate the full path integral on the relevant Lefschetz thimbles in order to bound the contributions of other critical points.

\begin{figure}
    \centering
    \includegraphics[scale=.7]{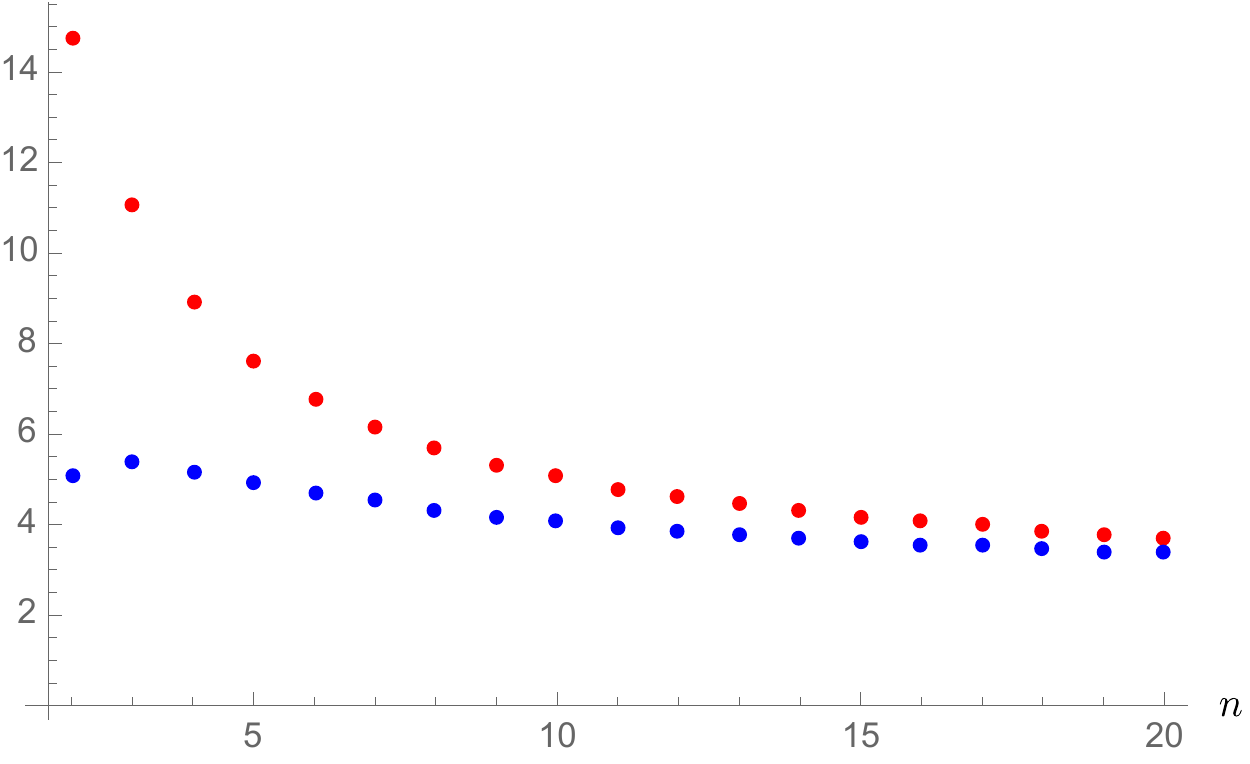}
    \caption{\small{The original volume conjecture sequence (left hand side of~\eqref{eq:vc}, blue) for the figure-eight knot, and the modified sequence (left hand side of~\eqref{eq:monotone-vc}, red). The non-monotonicity at the beginning of the original sequence is repaired by~\eqref{eq:monotone-vc}.  Both sequences are computed using Habiro's formula for the colored Jones polynomial of the figure-eight knot~\cite{Habiro:2000}.
    }}
    \label{fig:monotone-4_1}
\end{figure}

We observed in Section~\ref{sec:cs-implications} that there could be large discrepancies between the value of $\gamma$ for the early terms in the volume conjecture and the value of $\gamma$ at which the geometric conjugate connection is added to the path integral.
This motivates a new style of volume conjecture which could be engineered to be immediately monotonic.
We simply keep $\gamma \approx 1$ throughout the limit, though this must be done to carefully avoid integer level.
By avoiding integer level, we mean that, \textit{e.g.}, for $n=2$, if we simply solve $\gamma = 1$ we find $k=1$, and we already argued why this evaluation should yield no nontrivial information about the volume.
So we must instead begin at some value like $\gamma = 2$, which would correspond to $k = \frac{1}{2}$.
Moreover, we should tune this value with $n$ so that we approach $\gamma = 1$ in the semiclassical limit.
All these constraints lead to a version of the volume conjecture where we evaluate the relevant path integrals at a candidate value of $\gamma$ such as
\begin{equation}
    \gamma = \frac{n}{n-1} ~.
\label{eq:monotone-gamma}
\end{equation}
This corresponds to level
\begin{equation}
    k = \frac{(n-1)^2}{n} ~,
\end{equation}
which is certainly always fractional for integer $n\geq 2$ since the parities of the numerator and denominator do not match.
With this choice, a monotonic version of the volume conjecture would be (recalling that the prefactor $\frac{2\pi}{n}$ in the usual volume conjecture is really $\frac{2\pi}{k}$ from the Chern--Simons perspective)
\begin{equation}
    \lim_{n\to \infty} \frac{2\pi n \log |J_n(K;e^{2\pi i n / (n^2+1)})|}{(n-1)^2} = \text{Vol}(S^3 \setminus K) ~.
\label{eq:monotone-vc}
\end{equation}
This conjecture repairs the non-monotonicity in the volume conjecture sequence for the figure-eight knot (Figure~\ref{fig:monotone-4_1}).
Of course, we are not guaranteed that all knots still receive a contribution from the geometric conjugate connection at $\gamma = 2$.
Indeed, unfortunately,~\eqref{eq:monotone-vc} just barely fails to be monotonic for the knot $7_2$.
To formulate a more promising conjecture, we should assume that there is some finite radius around $\gamma=1$ where all knots still receive such a contribution, and begin the sequence there, which would lead to a different functional dependence than~\eqref{eq:monotone-gamma}.
As noted in~\cite{Garoufalidis_2005}, eventual monotonicity is important in proving that the limit in the volume conjecture actually exists, so it may be easier to prove that the limit exists in some improved version of~\eqref{eq:monotone-vc}. 

A less technically complex but more computationally intensive direction involves repeating our machine learning analysis for the $n>2$ colored Jones polynomials.
We expect to find a similar approximation function, perhaps with different coefficients, which converges quickly to the function which appears in the volume conjecture as $n$ increases.
While we attempted (and failed) to formulate a monotonic version of the usual volume conjecture above, there is a more general (and vague) family of conjectures which follow more directly from our numerics:
\begin{equation}
    a_n \log (|J_n(K;e^{2\pi i \gamma_n / (n+2\gamma_n-1)})|+b_n) + c_n \approx \text{Vol}(S^3 \setminus K) ~.
\label{eq:conjecture}
\end{equation}
The $\approx$ in the above expression means there is a margin for error which monotonically decreases to zero as $n \to \infty$, and furthermore we must have $a_n \to \frac{2\pi}{n}$, $b_n \to 0$, $c_n \to 0$, and $\gamma_n \to 1$ in this limit.
We view $a_n - \frac{2\pi}{n}$, $b_n$, and $c_n$ as ``quantum corrections'' which account for the large quantum effects in the path integral far from semiclassicality.
The quantity $\gamma_n$ begins around $\gamma_2 = \frac{3}{2}$, according to our numerical results in this work.
However, we leave unspecified its precise functional dependence on $n$, and similarly for the coefficients $a_n$, $b_n$, and $c_n$.
The main advantage of~\eqref{eq:conjecture} over the simpler~\eqref{eq:monotone-vc} is that we expect~\eqref{eq:conjecture} to already be roughly $97\%$ accurate immediately at $n=2$, whereas for~\eqref{eq:monotone-vc} we do not have such a guarantee.
Indeed, we expect that the error in~\eqref{eq:conjecture} is actually bounded, whereas~\eqref{eq:monotone-vc} can be arbitrarily wrong at small $n$, though an improved version of~\eqref{eq:monotone-vc} would still converge monotonically in $n$.
Of course, the price we pay for this immediate accuracy is the introduction of many free parameters in the conjecture.

\section*{Acknowledgments}
We are grateful to Onkar Parrikar for prior collaboration in machine learning aspects of knot theory.
Figure~\ref{fig:j2j3} was obtained in this joint work.
We thank Nathan Dunfield, Sergei Gukov, 
Yang-Hui He, Onkar Parrikar, and Edward Witten for discussions and comments on the manuscript.
We thank Dror Bar-Natan for computing HOMFLY-PT polynomials at $15$ crossings and for the \texttt{dbnsymb} \LaTeX\ symbol package~\cite{dbnsymb}.
JC and VJ are supported by the South African Research Chairs Initiative of the Department of Science and Technology and the National Research Foundation and VJ by a Simons Foundation Mathematics and Physical Sciences Targeted Grant, 509116.
AK is supported by the Simons Foundation through the It from Qubit Collaboration and thanks the University of Pennsylvania, where this work was initiated.

\appendix

\section{Basic neural network setup}\label{sec:nnsetup}
The neural networks used in these experiments were built in \texttt{Python 3}, using \texttt{GPU-Tensorflow} with a \texttt{Keras} wrapper~\cite{tensorflow2015-whitepaper}. The neural network has two hidden layers, $100$ neurons wide, with ReLU activation functions. The training used an \texttt{Adam} optimizer and a mean squared loss function. There were typically $100$ training epochs. The network was trained on $10\%$ of the $313,209$ knots up to (and including) $15$ crossings and tested on the remaining $90\%$. The code snippet below summarizes the key steps of building and training the neural network. 

\begin{lstlisting}[language=Python]
    import numpy as np
    import tensorflow as tf
    from sklearn.model_selection import train_test_split
    
    train_jones, test_jones, train_vols, test_vols = 
    train_test_split(jones, vols, test_size=0.9)
    
    model = tf.keras.models.Sequential([
    tf.keras.layers.Dense(100, activation='relu',
        input_dim=len(train_jones[0])),
    tf.keras.layers.Dense(100, activation='relu'),
    tf.keras.layers.Dense(1)
    ])
    
    model.compile(optimizer='adam', loss='mean_squared_error')
    
    model.fit(train_jones, train_vols, epochs=100, 
        batch_size=32, verbose=1)
\end{lstlisting}

\begin{figure}[t]
    \centering
    \includegraphics[scale=0.6]{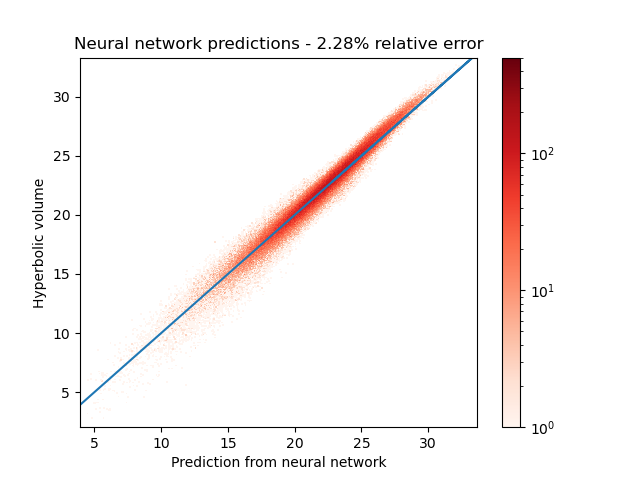}
    \caption{\small{Results for training the baseline neural network in \texttt{Python}. The network has two hidden layers, $100$ neurons wide, with ReLU activation functions.  }}
    \label{fig:basic-nn}
\end{figure}

\paragraph{Results:} Throughout the experiments, different representations/parts of the Jones polynomial were used as input to the neural network. To demonstrate the basic setup, the input data consists of $16$-vectors containing the coefficients of the Jones polynomials. The network generally achieves around $2.3\%$ relative error, as shown in Figure~\ref{fig:basic-nn}.

\section{Khovanov homology, Jones coefficients, and volume}\label{sec:khovanov}

\subsection{Khovanov homology}\label{sec:khovanov1}

Machine learning has been applied to string theory for a variety of purposes, beginning in~\cite{He:2017aed, Ruehle:2017mzq,Krefl:2017yox,Carifio:2017bov}.
Many instances in which machine learning has been successfully applied to mathematical physics involve an underlying homological structure~\cite{Bull:2018uow,Bull:2019cij,He:2020fdg,He:2020lbz}.\footnote{Such structures appear in string theory fairly often, and more sophisticated machine learning (beyond simple feed-forward network architecture) has also been applied fruitfully to the study of the string landscape.  For a small sample, see~\cite{Carifio:2017bov,Halverson:2019tkf,Halverson:2019vmd,Halverson:2020opj,Bies:2020gvf}, and also see the review~\cite{RUEHLE20201} for more complete references in this area.}
In certain cases, analytic formulas inspired by machine learning results have been obtained~\cite{Brodie:2019dfx,Brodie:2019ozt,Brodie:2019pnz}.
Knot theory appears to be no exception to this pattern, as there is an underlying homology theory related to the Jones polynomial known as Khovanov homology~\cite{khovanov2000}.
As we will make reference to the homological nature of the Jones polynomial at several points in this supplemental discussion, here we provide a brief explanation of Khovanov homology following the compact treatment in~\cite{Bar_Natan_2002}.

The Jones polynomial, in its form more familiar to mathematicians, is a Laurent polynomial in a variable $q$ defined by a skein relation\footnote{Knot theorists will notice that this is a rather non-standard skein relation.  We will comment on this, and related normalization issues, in Appendix~\ref{sec:norm}.}
\begin{equation}
\includegraphics[scale=.55]{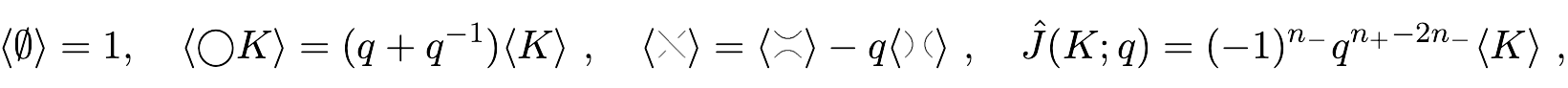}
\label{eq:skein}
\end{equation}
where $\avg{K}$ is known as the Kauffman bracket~\cite{HKAUFFMAN1987395} of the knot $K$, and $n_\pm$ are the number of right- and left-handed crossings in a diagram of $K$.
The quantity $n_+ - n_-$ is often called the writhe of the knot diagram.
We have used the notation $\hat{J}$ because this object is not quite equivalent to our $J_2(K;q)$.
It is unnormalized in the following sense:
\begin{equation}
    \hat{J}(K;q) = (q+q^{-1}) J_2(K;q^2) ~.
\label{eq:unnormalized-jones}
\end{equation}
The skein relation for the Kauffman bracket involves two ``smoothings'' of a given crossing.
Of course, in each of these smoothings, the total number of crossings is reduced by one.
The recursion terminates when all crossings have been smoothed, so the set of binary strings of length $c$, $\{0,1\}^c$, is the set of total smoothings for a knot diagram with $c$ crossings.

Khovanov's homology, roughly speaking, begins by assigning a vector space $V_\alpha(K)$ to each string $\alpha \in \{0,1\}^c$.
If $V$ is the two-dimensional graded vector space with basis elements $v_\pm$ and $\deg (v_\pm) = \pm 1$, then $V_\alpha(K) \equiv V^{\otimes k}\{r\}$ where the total smoothing $\alpha$ of $K$ results in $k$ closed loops, and $\{r\}$ is the degree shift by $r$.\footnote{Recall that a graded vector space $W = \oplus_m W_m$ has graded dimension $q\dim W = \sum_m q^m \dim W_m$.  Furthermore, the degree shift produces another graded vector space $W\{r\}$ with homogeneous subspaces $W\{r\}_m \equiv W_{m-r}$, so $q\dim W\{r\} = \sum_m q^{m+r} \dim W_m$.}
The height of a string $\alpha$ is defined as the number of ones, $|\alpha| = \sum_i \alpha_i$.
The strings with equal height $r$ can be grouped together, and (through some more work) the corresponding vector spaces can be assembled into a doubly-graded homology theory $\mathcal{H}^r(K)$, with Poincar\'{e} polynomial
\begin{equation}
    Kh(K; t,q) \equiv \sum_r t^r q\dim \mathcal{H}^r(K) ~. 
\end{equation}
Khovanov proved that this homology theory is a knot invariant, and that its graded Euler characteristic is the unnormalized Jones polynomial
\begin{equation}
    \chi_q(K) = Kh(K;-1,q) = \hat{J}(K;q) ~.
\end{equation}
Another useful quantity associated with Khovanov homology is the Khovanov rank, given by
\begin{equation}
    \text{rank}(\mathcal{H}^\bullet(K)) = \sum_r q\dim \mathcal{H}^r(K) ~.
\end{equation}
The Khovanov rank is correlated with the hyperbolic volume, as noticed in~\cite{khovanov2003}, and was compared to neural network prediction techniques in~\cite{Jejjala:2019kio}.

The main point we wish to convey with this brief summary is that the coefficients of the Jones polynomial are related to dimensions of homology groups, and therefore any success of our machine learning techniques represents another piece of evidence that underlying homological structure is relevant for characterizing physics and mathematics which is machine-learnable.

\subsection{Power law behavior of coefficients}
\begin{figure}[t]
   \centering
\begin{tabular}{cccc}
\includegraphics[width=3cm]{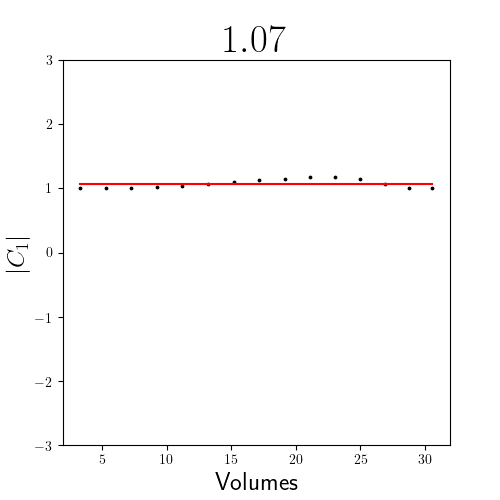}&
\includegraphics[width=3cm]{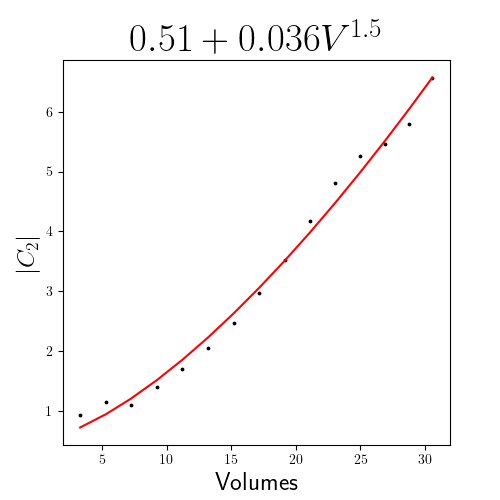}&
\includegraphics[width=3cm]{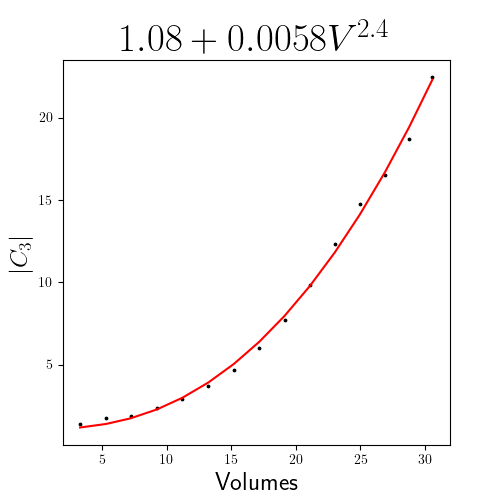}&
\includegraphics[width=3cm]{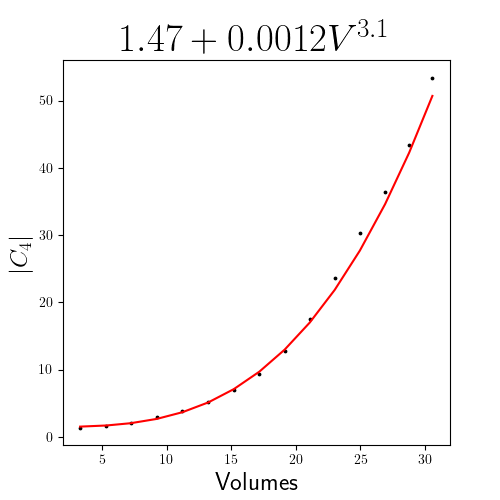}\\
\includegraphics[width=3cm]{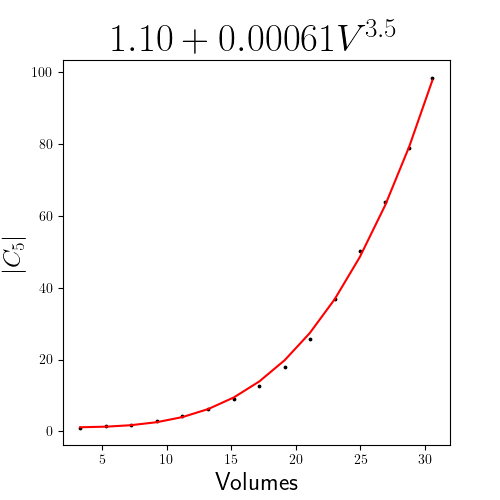}&
\includegraphics[width=3cm]{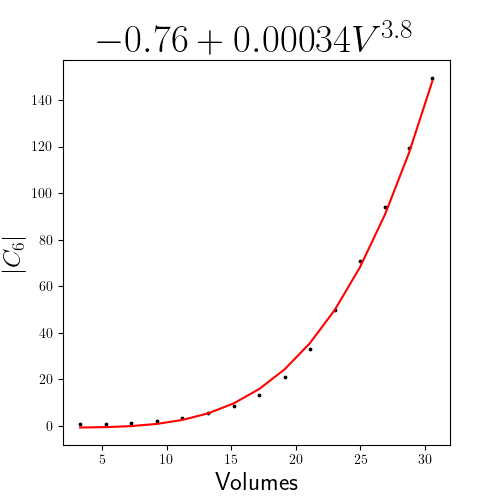}&
\includegraphics[width=3cm]{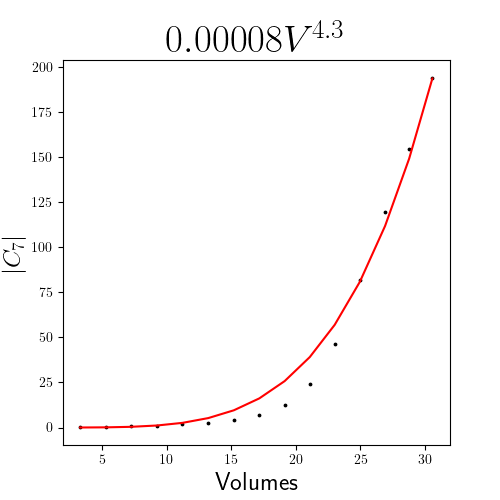}&
\includegraphics[width=3cm]{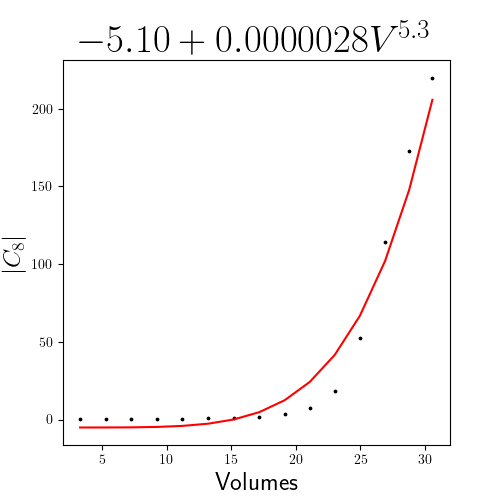}
\end{tabular}
    \caption{\small{The absolute values of the (first $8$) Jones polynomial coefficients as a function of the hyperbolic volume. The coefficients obey a power law relationship with respect to the volume, where (for $c_i$ the $i$\textsuperscript{th} coefficient) the constant coefficient decreases (by about an order of magnitude) and the exponent increases as $i$ increases. These power law relationships are symmetric with the last $8$ coefficients, as demonstrated in the Table~\ref{tab:coefs}.}}
\label{fig:coefs}
\end{figure}

In this appendix, we study how the coefficients in the Jones polynomial may be individually related to the hyperbolic volume. 
Through Khovanov homology, we know that each coefficient is measuring the dimension of a certain homology group.
Thus, this numerical study is in some sense a generalization of Khovanov's study of patterns in his homology~\cite{khovanov2003}, and his observation that coarser properties of his homology are related to the volume (see~\cite{Jejjala:2019kio} for more explanation and analysis of this point).
Patterns in Jones polynomial coefficients were also studied in~\cite{Walsh2014,Walsh2018}. 

The vectors of coefficients is represented as $\mathbf{J} = ( c_1,  \ldots,  c_8,  c_{-8},  \ldots,  c_{-1}),  
$ where padding with zeros is done from the center so that, say, $c_1$ is always the first nonzero coefficient in the polynomial and $c_{-1}$ is always the last nonzero coefficient in the polynomial. The volume is binned into $16$ intervals and the mean absolute values of each coefficient is calculated in the bins. Using \texttt{Mathematica} curve fitting procedures, it was found that the (absolute) coefficients obey power law relationships with respect to the volume; $|c_i| \rightarrow a V ^ b$. For $i \in [1, 8]$, we find that $c_i$ and $c_{-i}$ always obey (nearly) the same power law relationship. Additionally, as $i$ increases, the exponent $b$ increases and $a$ decreases by roughly an order of magnitude. These relationships are given in Table~\ref{tab:coefs} and Figure~\ref{fig:coefs}.

\begin{table}
\begin{center}
    \begin{tabular}{|c|c|c|c|c|}
        \hline
        Coefficient & Function & Constant term & $V$ exponent & $V$ coefficient\\
        \hline
        $c_1$ & $1.0662$ & $1.0662$ & -- & -- \\
        \hline
        $c_{-1}$ & $1.07462$ & $1.07462$ & -- & --\\
        \hline
        \hline
        $c_2$ & $0.506466 + 0.035896V^{1.5}$ & $0.506466$ & $1.5$ & $0.035896$ \\
        \hline
        $c_{-2}$ & $0.709162 + 0.036988 V^{1.5}$ & $0.709162$ & $1.5$ & $0.0369388$\\
        \hline
        \hline
        $c_3$ & $1.0797 + 0.00578349V^{2.4}$ & $1.0797$ & $2.4$ & $0.00578349$ \\
        \hline
        $c_{-3}$ & $1.1049 + 0.00607516 V^{2.4}$ & $1.1049$ & $2.4$ & $0.00607516$\\
        \hline
        \hline
        $c_4$ & $1.46782 + 0.0012235V^{3.1}$ & $1.46782$ & $3.1$ & $0.0012235$ \\
        \hline
        $c_{-4}$ & $1.26703 + 0.00188961 V^3$ & $1.26703$ & $3$ & $0.00188961$\\
        \hline
        \hline
        $c_5$ & $1.1018 + 0.0006121V^{3.5}$ & $1.1081$ & $3.5$ & $0.0006121$\\
        \hline
        $c_{-5}$ & $0.000629931V^{3.5}$ & $0$ & $3.5$ & $0.000629931$\\
        \hline
        \hline
        $c_{6}$ & $-0.758 +0.000338V^{3.8}$ & $-0.758$ & $3.8$ & $0.000338$\\
        \hline
        $c_{-6}$ & $0.0003446V^{3.8}$ & $0$ & $3.8$ & $0.000344$\\
        \hline
        \hline
        $c_{7}$ & $0.0000795V^{4.3}$ & $0$ & $4.3$ & $0.0000795$\\
        \hline
        $c_{-7}$ & $0.0000803V^{4.3}$ & $0$ & $4.3$ & $0.0000803$\\
        \hline
        \hline
        $c_{8}$ & $-5.104 + 0.00000283V^{5.3}$ & $-5.104$ & $5.3$ & $0.00000283$\\
        \hline
        $c_{-8}$ & $-5.0795 + 0.00000283V^{5.3}$ & $-5.079$ & $5.3$ & $0.00000283$\\
        \hline
    \end{tabular}
\end{center}
    \caption{\small{Fits for the mean absolute value of the coefficients as a function of the volumes.}}
\label{tab:coefs}
\end{table}

\section{Normalizations, orientation, and framing}\label{sec:norm}

The precise definition of the Jones polynomial can sometimes differ between physicists and mathematicians.
Fundamentally, the path integrals which enter~\eqref{eq:math-defn} that we have taken to define $J_n(K;q)$ produce invariants of an oriented, framed link in $S^3$.
Actually, since reversing the orientation of a Wilson loop corresponds to conjugation of the associated representation, and representations of $SU(2)$ are real, the $SU(2)$ Chern--Simons path integral defines an invariant of unoriented, framed links in $S^3$.
The writhe factor in~\eqref{eq:skein} serves to exchange the dependence on framing of the Kauffman bracket for a dependence only on orientation; it is a sort of normalization which removes the extra factors of $q^{3/4}$ that may be introduced by type I (unframed) Reidemeister moves.\footnote{See the discussion around Corollary 2.8 and Remark 2.9 in~\cite{Le2014}.}
This is more or less the same as selecting the 0-framing for all components of a link.
So, the path integral formalism incorporates framing changes naturally, whereas the values for the Jones polynomial often quoted in the mathematical literature effectively assume a fixed framing which can always be selected for links in $S^3$.

An additional detail arises when we make a more direct comparison between the path integrals found in~\eqref{eq:math-defn} and the skein relation definition of the Jones polynomial.
To begin, we refer to~\cite{Witten:1988hf}, where a skein relation was derived using the path integral directly.
In equations (4.22) and (4.23) of~\cite{Witten:1988hf}, the $SU(2)$ skein relation for the path integral was found to be
\begin{equation}
    q Z(L_+) - q^{-1} Z(L_-) = (q^{1/2}-q^{-1/2}) Z(L_0) ~,
\label{eq:skein-witten}
\end{equation}
where $L_+$, $L_-$, and $L_0$ represent the same link but with a single crossing modified (if necessary) to be an over-crossing, under-crossing, or uncrossed (away from this crossing, the link is unchanged).
Pictorially,~\eqref{eq:skein-witten} is
\begin{equation}
\includegraphics[scale=.55]{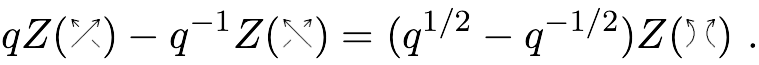}
\end{equation}
Other than this relation, the only other piece of information needed to determine the Jones polynomial (as defined in~\eqref{eq:math-defn}) is the expectation of the unknot.
If we normalize the path integral as in~\eqref{eq:math-defn}, the unknot has expectation $1$.
\begin{equation}
    J_2(0_1;q) = 1 ~.
\end{equation}
This is equivalent to requiring that the Jones invariants should be multiplicative under connected sum of links, $J_2(L_1 \# L_2;q) = J_2(L_1;q) J_2(L_2;q)$.
The skein relation~\eqref{eq:skein-witten} actually matches exactly with the one written above Remark 2.9 in~\cite{Le2014} for the writhe-normalized Jones polynomial $\overset{\circ}{V}_L$ (in the notation of~\cite{Le2014}).\footnote{There is a typo in equation (6) of~\cite{Le2014}.  The coefficient on $\overset{\circ}{V}_{L_-}$ should be $q^{-1}$.  The correct expression can be derived by multiplying equation (3) of~\cite{Le2014} by $q^{-(3/4)w(L_0)}$ and rewriting the coefficients of ${V}_{L_+}$ and ${V}_{L_-}$ in terms of $q^{-(3/4)w(L_+)}$ and $q^{-(3/4)w(L_-)}$, respectively.  To do so, observe that the writhe obeys $w(L_\pm) = w(L_0) \pm 1$.}
Thus, the ratio in~\eqref{eq:math-defn} and $\overset{\circ}{V}_L/\overset{\circ}{V}_{0_1}$ from~\cite{Le2014} will be equivalent up to a power of $q^{3/4}$ associated with a change of framing in the numerator of~\eqref{eq:math-defn}.

However, other formulations of the skein relation favored by some mathematicians, such as the one used in~\cite{Jones2014,Jones2014long}, differ from~\eqref{eq:skein-witten}.
The skein relation in~\cite{Jones2014} is
\begin{equation}
    q^{-1} V_{L_+} - q V_{L_-} = (q^{1/2} - q^{-1/2}) V_{L_0} ~.
\label{eq:skein-jones}
\end{equation}
It is clear that~\eqref{eq:skein-jones} is not equivalent to~\eqref{eq:skein-witten}, so which of these determines the ``true'' Jones polynomial?\footnote{The relationship between these two skein relations is likely well-known to knot theorists, but we have not been able to find a discussion in the literature, so we provide a self-contained treatment in what follows.}
This difference is not a question of normalization, because the skein relations are all linear, so any normalization obeys the same relation.
We will see that the two skein relations are related by a simple, but non-obvious, transformation.
Let $L'$ be the mirror of $L$; in other words, $L$ with all over-crossings changed to under-crossings and vice-versa.
Also, let $|L|$ be the number of disjoint circles in $L$.
Then, we have
\begin{equation}
    V_{L'}(q) = (-1)^{|L|+1} J_2(L;q) ~.
\label{eq:jones-witten-relation}
\end{equation}
We begin by noting that $V_{L'}$ obeys a skein relation which is derived by simply switching $L_+$ and $L_-$ in~\eqref{eq:skein-jones}, since this is the effect of the mirroring operation on a given crossing.
\begin{equation}
    q^{-1} V_{L_-'} - q V_{L_+'} = (q^{1/2} - q^{-1/2}) V_{L_0'} ~.
\label{eq:skein-mirror}
\end{equation}
We now observe a few relationships among $|L_+|$, $|L_-|$, and $|L_0|$.
Since moving one strand through another does not change the number of components, $|L_+| = |L_-|$.
Furthermore, if the crossing in $L_\pm$ is between two different components, then $|L_\pm| = |L_0| + 1$, as in $L_0$ these two components are joined.
Similarly, if the crossing in $L_\pm$ is only one component crossing itself, then $|L_\pm| = |L_0| - 1$, as in $L_0$ this component is split into two.
The upshot is that $|L_0|$ always has opposite parity from $|L_\pm|$, so
\begin{equation}
    (-1)^{|L_+|} = (-1)^{|L_-|} = (-1)^{|L_0|+1} ~.
\label{eq:component-parity}
\end{equation}
Dividing~\eqref{eq:skein-witten} by $Z(0_1;k)$ produces an identical skein relation for $J_2$, and multiplying by $(-1)^{|L_+|+1}$ and using~\eqref{eq:component-parity} yields
\begin{equation}
    q (-1)^{|L_+|+1} J_2(L_+) - q^{-1} (-1)^{|L_-|+1} J_2(L_-) = -(q^{1/2}-q^{-1/2}) (-1)^{|L_0|+1} J_2(L_0) ~.
\end{equation}
Interpreted as a skein relation for $(-1)^{|L|+1} J_2(L)$, this equation is equivalent to~\eqref{eq:skein-mirror}.
Therefore, the quantity $(-1)^{|L|+1} J_2(L)$ obeys the same skein relation as $V_{L'}$, and since they take the same initial value on the unknot, they are equal, proving~\eqref{eq:jones-witten-relation}.

The Jones polynomial is not invariant under the mirroring operation.
However, it changes in a predictable way.
Since the transformation $q \to q^{-1}$ changes~\eqref{eq:skein-jones} into~\eqref{eq:skein-mirror}, we have
\begin{equation}
    V_{L'}(q) = V_L(q^{-1}) ~.
\end{equation}
For the purposes of the volume conjecture, however, it does not matter whether we work with $L$ or $L'$.
This is because $S^3 \setminus L$ is homeomorphic to $S^3 \setminus L'$, and since the hyperbolic volume is a topological invariant by the Mostow--Prasad rigidity theorem, their volumes are equal.
Furthermore, the evaluations of the colored Jones invariants which appear in the volume conjecture are simply complex conjugated under $q \to q^{-1}$ because $q$ is a root of unity, which leaves unchanged the magnitude $|J_n(e^{2\pi i /n})|$.
That being said, when we write numerical results for~\eqref{eq:math-defn}, we ought to keep in mind that (for example) the right-handed trefoil knot (three over-crossings) obeys
\begin{equation}
    J_2(3_1;q) = q^{-1} + q^{-3} - q^{-4} = V_{3_1}(q^{-1}) ~.
\end{equation}
So, if our dataset consists of left-handed knots with polynomials derived using~\eqref{eq:skein-jones}, we will write explicit evaluations of~\eqref{eq:math-defn} for the corresponding right-handed knots.
Interestingly, certain generalizations of the basic volume conjecture~\eqref{eq:vc} (including the ``complex volume conjecture'' which involves the Chern--Simons invariant CS$(S^3\setminus K)$) are not invariant under an exchange of handedness on one side, which means that whenever one writes something like the complex volume conjecture~\cite{murakami2002kashaev}
\begin{equation}
    \lim_{n\to\infty} \frac{2\pi \log J_n(K;e^{2\pi i /n})}{n} = \text{Vol}(S^3\setminus K) + 2\pi^2 i \text{CS}(S^3 \setminus K) ~,
\label{eq:complex-vc}
\end{equation}
there is actually a unique skein relation which defines $J_n$ in such a way that~\eqref{eq:complex-vc} could be true.
We have not seen a discussion in the literature of this point, and it would be interesting to determine which of~\eqref{eq:skein-jones} and~\eqref{eq:skein-witten} is the correct choice.
We suspect the correct relation is the one which comes from Chern--Simons theory,~\eqref{eq:skein-witten}.
If we were instead to use the alternate skein relation, we would need to take the complex conjugate of the right hand side in~\eqref{eq:complex-vc} to have a chance of being correct.

We will explain one final issue related to normalizations.
The $q^2$ in the argument of $J_2$ on the right hand side of~\eqref{eq:unnormalized-jones} arises because the skein relation we have written in~\eqref{eq:skein} matches the rather non-standard choice in~\cite{Bar_Natan_2002}, which in turn matches~\cite{khovanov2000}. 
Unlike the more standard relations discussed in~\cite{Jones2014,Le2014} that we addressed above,~\eqref{eq:skein} has the advantage of generating a Laurent polynomial $\hat{J}(K;q)$ on links with any number of components.
Therefore, it differs in detail from the path integral ratio~\eqref{eq:math-defn}, which may require a change of framing to produce a Laurent polynomial.
As a consequence of these details, the relationship~\eqref{eq:unnormalized-jones} should be understood up to a change of framing of the right hand side.
Since we will always consider Laurent polynomials in this work, we assume a framing where the left hand side of~\eqref{eq:math-defn} is a Laurent polynomial that matches the output of the skein relation in~\cite{Jones2014}.

\section{t-distributed stochastic neighbor embedding of knot invariants}\label{sec:tsne}

\subsection{t-distributed stochastic neighbor embedding}

t-distributed stochastic neighbor embedding (t-SNE) is a method used to visualize higher-dimensional data in two or three dimensions~\cite{vanDerMaaten2008}. It allows one to get a visual understanding of the underlying structure of the data. t-SNE works by preserving the distances between the data points. By centering a Gaussian over the point $x_i$ and measuring the density of all the points $j$ under that Gaussian, the quantity $p_{j|i}$ is calculated. This can be viewed as the conditional probability that $i$ would pick $j$ as a nearest neighbor. In the lower-dimensional space, the analogous quantity $q_{j|i}$ is calculated. Both of these values are symmetrized ($p_{ij} = \frac{1}{2}(p_{i|j} + p_{j|i})$). To ensure that $q_{ij}$ accurately represents $p_{ij}$, the sum of Kullback--Leibler divergences (relative entropies) over all data points, 
\begin{equation} 
D_{\text{KL}}(p \| q) = \sum_{i}\sum_{j \neq i} p_{ij}\log{\frac{p_{ij}}{q_{ij}}} ~,
\end{equation}
is minimized.
Thus, the minimization of this quantity with respect to $q$ determines a map from the high dimensional space to the lower dimensional space, which can then be used to project the input.
An important parameter which we vary in our application is the perplexity, which quantifies the width of the Gaussians around each data point in terms of the local data point density.
Greater perplexity roughly corresponds to considering more points as ``neighbors", and extends the Gaussian widths.
In this way, the t-SNE algorithm is a sort of constrained optimization of $q$, and robust results should not depend strongly on the precise form of this constraint.

\subsection{Volume and Chern--Simons invariant}\label{sec:vcsi}
\begin{figure}[ht]
\centering
    \includegraphics[scale=0.4]{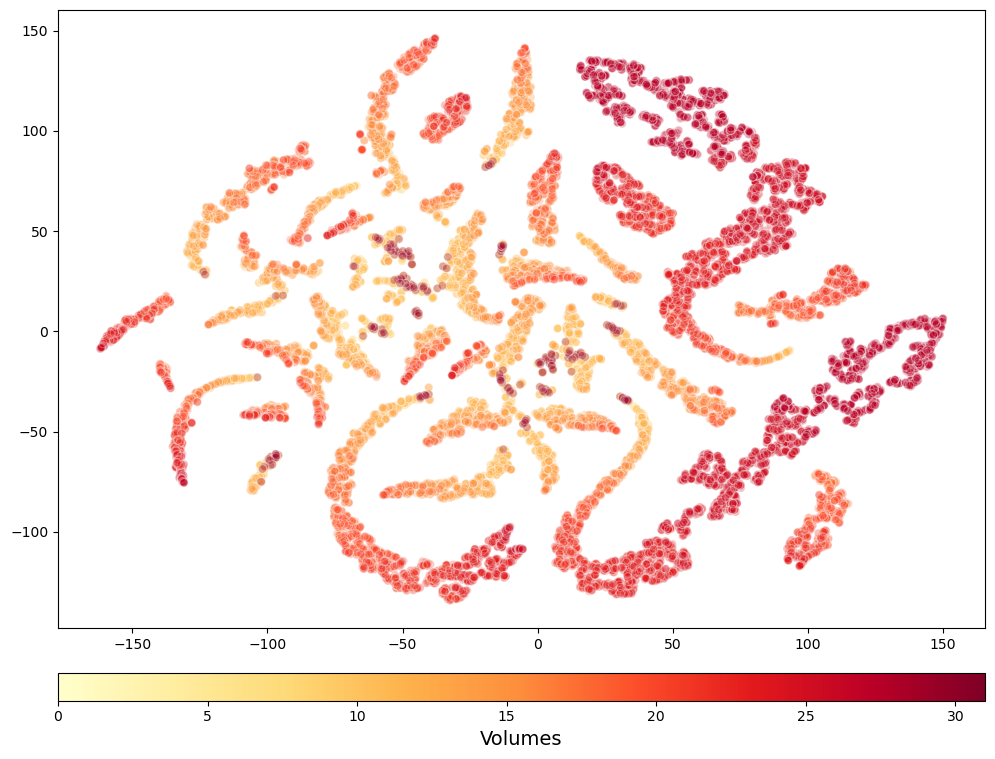}
    \caption{\small{t-SNE plot of 2000 Jones polynomials, colored by volume (default perplexity of $30$).}}
    \label{fig:tsne}
\end{figure}

We used t-SNE to reduce the input vector of Jones polynomial coefficients to a two-dimensional space for visualization purposes (Figure~\ref{fig:tsne}).
The t-SNE analysis was performed using \texttt{Python}'s \texttt{scikit-learn} library~\cite{scikit-learn}.
There are two key things to notice in this visualisation. 
First, the points clearly separate into clusters - this indicates that there is some clustering or hidden structure that separates knots by their Jones polynomial.
While this alone could be considered an artifact of the algorithm, we next notice that the points are colored by volume and it is clear that similar colors tend to cluster together.
The results here are fairly robust under changes in the internal parameters of the algorithm, like the perplexity, which indicates that the clustering effects we see really do exist in our data.

\begin{figure}[ht]
\centering
    \includegraphics[scale=0.4]{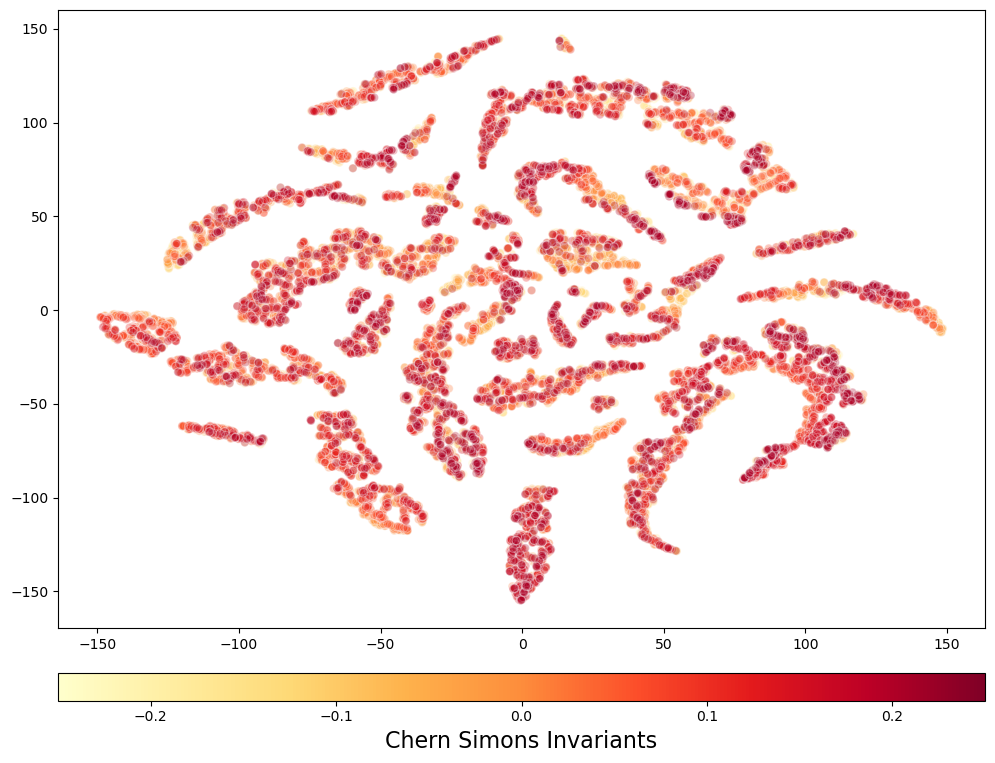}
    \caption{\small{t-SNE plot of 2000 Jones polynomials, colored by Chern--Simons invariant (default perplexity of $30$).}}
    \label{fig:tsne_cs}
\end{figure}

We emphasize that, even though LRP played the largest r\^ole in our analysis of the neural network, t-SNE is a purely data-driven technique which is extremely useful for building intuition about which quantities may be machine learnable from others. For instance, Figure~\ref{fig:tsne} demonstrates noticeable clustering, while Figure~\ref{fig:tsne_cs} does not. This is evidence that the hyperbolic volume may be machine learnable from the Jones polynomials, while the Chern--Simons invariant may not be.
Indeed, this intuition is confirmed in our experiments: none of the neural networks we employed in this work were able to reliably predict the Chern--Simons invariant of a knot from its Jones polynomial.

As a more general lesson, through the creation of plots like Figures~\ref{fig:tsne} and~\ref{fig:tsne_cs}, we can visualize data without any choice of learning framework.
This technique may have promising applications in other areas of experimental mathematics or physics-inspired data science where machine learning sometimes involves prohibitively large computational resources.
In those cases, t-SNE could be used to quickly iterate through hypotheses about correlations in the dataset without running a full learning algorithm.

\section{Other experiments}\label{sec:more_exp}

In this appendix, we state the results of a number of mathematical experiments.
In applying artificial intelligence to problems in theoretical physics and mathematics, there is no \textit{a priori} understanding of what features are machine learnable.
The problems for which machine learning provides a useful prediction are often discovered by trial and error.
Having isolated such an instance in connecting the Jones polynomial with the hyperbolic volume, we believe it is useful to publicize variations on this central theme.

We as well note some of the less successful experiments among our investigations.
When reverse engineering a neural network to obtain analytic results, we expect failed experiments to also guide us.
As reported in~\cite{Jejjala:2019kio} and in Appendix~\ref{sec:vcsi}, the fact that the volume of the knot complement and the Chern--Simons invariant appear on an equal footing in the generalized volume conjecture~\eqref{eq:complex-vc} but only the former is machine learnable from $J_2(K;q)$ using a simple neural network architecture is particularly striking and perhaps worth highlighting again.
It should also be noted that using evaluations of $J_2(K;q)$ at complex phases, we were unable to predict the Chern--Simons invariant.

\subsection{The HOMFLY-PT and Khovanov polynomials}
The HOMFLY-PT polynomial $P(K;a,z)$~\cite{Freyd:1985dx,Przytycki:1987} is a two variable knot invariant that generalizes both the Alexander polynomial $\Delta(K;q)$~\cite{alexander1928topological} and the Jones polynomial $J_2(K;q)$.
In particular,
\begin{equation}
    \Delta(K;q) = P(K;1,q^{1/2}-q^{-1/2}) ~, \qquad
    J_2(K;q) = P(K;q^{-1},q^{1/2}-q^{-1/2}) ~.
\end{equation}
We have seen in Appendix~\ref{sec:khovanov1} that the Khovanov polynomial $Kh(K;t,q)$~\cite{khovanov2000,Bar_Natan_2002} is another such two variable polynomial invariant with the property that
\begin{equation}
    J_2(K;q^2) = \frac{Kh(K;-1,q)}{q+q^{-1}} ~.
\end{equation}
Using a similar architecture to what is quoted in Appendix~\ref{sec:nnsetup}, we machine learn the volume from the HOMFLY-PT polynomial to an accuracy of $93.9$\% from the full dataset of knots up to $15$ crossings.
The Khovanov polynomial predicts the volume of the knot complement to an accuracy of $97.2$\% from a partial dataset of $196,002$ knots up to $15$ crossings.
In both experiments, we use $10$\% of the dataset for training.
The input is a flattened matrix of coefficients appearing in the polynomials where the dimensions of the matrix are determined by the difference between the maximum and minimum degrees of the polynomials in each of the two variables.
We notice that though the HOMFLY-PT polynomial contains more information about the knot than the Jones polynomial, it performs significantly worse when used as the input in a two hidden layer neural network.

\subsection{Representations of $J_2(K;q)$}\label{sec:repofJ}
In~\cite{Jejjala:2019kio}, the Jones polynomial was faithfully represented as a vector.
The first two terms of the input vector are the minimum and maximum degree of the polynomial, and the subsequent entries are the coefficients in the polynomial, padded to the right with zeros to form an $18$-vector for knots up to $15$ crossings and a $19$-vector for knots up to $16$ crossings.
If we represent the Jones polynomial, again faithfully, as a sparse vector in a long representation whose length is determined by the difference between the maximum and minimum degrees of all polynomials in the dataset, the performance is unchanged.

Surprisingly, unfaithful representations of the Jones polynomial perform almost as well.
If we eliminate the degrees from the vector and simply train on the coefficients (that is to say, dimensions of particular Khovonov homology groups that differ for the various inputs), the mean absolute error increases only slightly, from $2.65$\% to $2.75$\%.
If the input vector sorts the coefficients from smallest to largest, the results remain largely unchanged, but if the ordering of the elements in each vector is randomized, then the mean absolute error is $6.44$\%, a significant loss of accuracy.
While the ordering of the coefficients does matter (or else the randomized vectors would perform just as well), the sorted vectors provide reasonable accuracy, so some rearrangements are allowed.
Linear combinations of the coefficients also give an error less than $3$\%.

Let us drop the degrees but preserve the ordering of coefficients.
This yields $16$-vector inputs for knots up to $15$ crossings.
In Appendix~\ref{sec:tsne}, t-SNE was used to reduce the data to two dimensions for visualization purposes: it allowed us to see the clustering of the inputs according to volumes.
Dimensional reduction may also be useful in determining how the neural network learns the volume.
Principal component analysis (PCA) is used as a method to reduce the dimension of the data (length $16$ to length $4$).
With this reduced input, the error grows to $2.80$\%.

Extending Dunfield's initial observation~\cite{Dunfield2000} that for alternating knots $\log|J_2(K;-1)|\propto \text{Vol}(S^3 \setminus K)$, we consider the Taylor expansion of $J_2(K;e^{i\pi(1-x)})$.
For instance, for the figure-eight knot $4_1$, we have
\begin{eqnarray}
    J_2(4_1;q) &=& q^{-2} - q^{-1} + 1 - q + q^2 ~, \\
    J_2(4_1;e^{i\pi(1-x)}) &=& 5 - \frac{10\pi^2}{2!} x^2 + \frac{34\pi^4}{4!} x^4 + \ldots ~.
\end{eqnarray}
We adopt the convention of dropping factors of $i$ and $\pi$ and write the first five coefficients as a vector; this is $(5,0,-5,0,1.417)$ in the example $4_1$.
Using a dataset of $10,000$ alternating knots between $11$ and $15$ crossings, the error in the neural network prediction of the volume was $6.5$\%, and the performance was particularly poor at low volumes.
Interesting, training on the first coefficient alone, \textit{i.e.}, training on just $J_2(K;-1)$ yields superior performance.

\subsection{Small networks}
\begin{figure}
    \centering
    \includegraphics[scale=0.45]{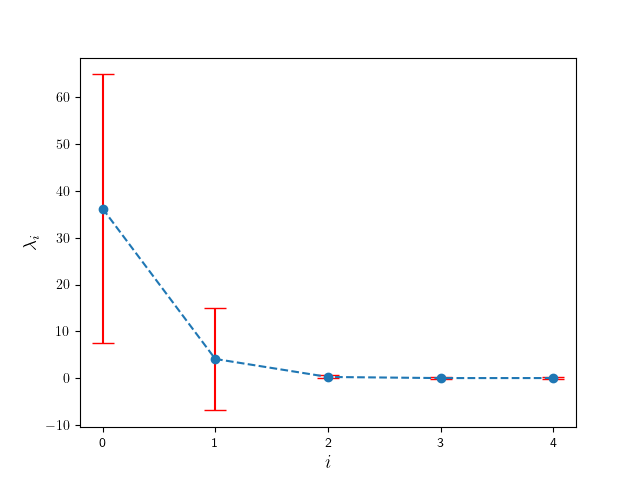}
    \includegraphics[scale=0.45]{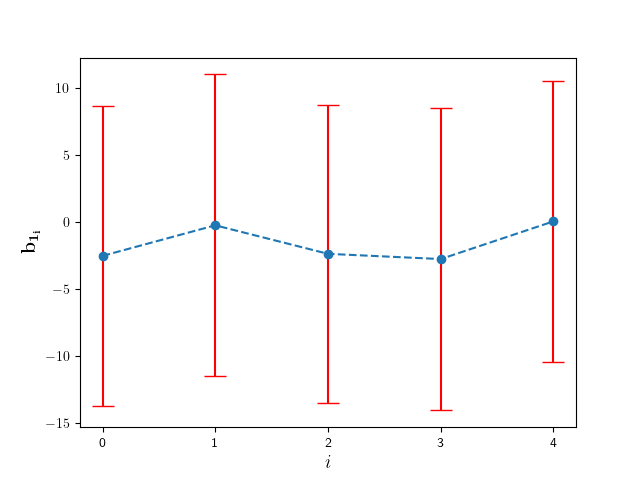}
    \caption{\small{(Left) Eigenvalues of the matrix $(W_1)^TW_1$, with $W_1$ being the weight matrix of the first layer, averaged over $100$ runs. (Right) The bias vector $\mathbf{b}_1$ of the first layer, averaged over $100$ runs. Error bars mark standard deviations.}}
    \label{fig:small_net_wb}
\end{figure}

In this experiment, we trained a neural network with one hidden layer, five neurons wide, using the relevant roots from the $r=8$ row of Table~\ref{tab:root_table} as input. Since $J_2(K;-1)$ is always real, the input vector is also of length $5$. There are $36$ parameters to be optimized: $25$ from the $5\times 5$ weight matrix after the input layer, $5$ from the bias vector added to the input layer, $5$ from the weight vector between the hidden layer and the output layer, and $1$ for the bias added before the output layer. Averaged over $100$ training runs on the same dataset, the network achieved an error of $5.68\%$. The large standard deviation on the bias vectors (Figure~\ref{fig:small_net_wb}, right) suggests that the bias vectors may be irrelevant in the function that the neural network is learning. Indeed, setting all five elements of the bias vector to zero does not significantly impact the performance of the network. Although three of the five eigenvalues are zero in Figure~\ref{fig:small_net_wb} (left), the performance is significantly worsened when fewer than five neurons are used.

The fact that such a simple neural network was able to perform so well is evidence that a simple analytic function exists between the Jones polynomial and the hyperbolic volume. The initial intention of this experiment was to write out the function learnt by the neural network and use it to develop a simple analytic function. However, the number of parameters (which is after zeroing out the bias vector in the first layer) as well as the application of the activation function made the neural network difficult to interpret, despite its relatively small size.

\subsection{Symbolic regression}\label{sec:symbolic}
In~\cite{cranmer2020discovering}, a combination of deep learning and symbolic regression is used to discover analytic expressions. A deep learning model, such as a Graph Neural Network, is trained on available data. Symbolic expressions are then fitted to the functions learned by the model. The software \texttt{PySR}~\cite{pysr} automates this process. We used \texttt{PySR} to look for functions of the form $V(K) = f(|J_2(K;e^{3\pi i/4})|)$, allowing the search to use all functions provided in the package. The best fit predicts the volumes to $96.56\%$ accuracy.  While the precise formula is unenlightening, it does include a logarithm of $J_2(K;e^{3\pi i/4})$, which reassures us that the formulas proposed in Section~\ref{sec:deep} are reasonable ans\"{a}tze.

\bibliographystyle{JHEP}
\bibliography{knots}

\end{document}